\documentclass[useAMS,usenatbib]{mn2e}
\input{epsf}
\usepackage{amssymb}
\usepackage{natbib}
\usepackage{color}
\usepackage{journals}
\usepackage{graphicx}

\newbox\grsign \setbox\grsign=\hbox{$>$} \newdimen\grdimen \grdimen=\ht\grsign
\newbox\simlessbox \newbox\simgreatbox
\setbox\simgreatbox=\hbox{\raise.5ex\hbox{$>$}\llap
     {\lower.5ex\hbox{$\sim$}}}\ht1=\grdimen\dp1=0pt
\setbox\simlessbox=\hbox{\raise.5ex\hbox{$<$}\llap 
     {\lower.5ex\hbox{$\sim$}}}\ht2=\grdimen\dp2=0pt

\newcommand{\hMpc}{{\ifmmode{h^{-1}{\rm Mpc}}\else{$h^{-1}$Mpc }\fi}}
\newcommand{\hGpc}{{\ifmmode{h^{-1}{\rm Gpc}}\else{$h^{-1}$Gpc }\fi}}
\newcommand{\hkpc}{{\ifmmode{h^{-1}{\rm kpc}}\else{$h^{-1}$kpc }\fi}}
\newcommand{\hMsun}{{\ifmmode{h^{-1}{\rm {M_{\odot}}}}\else{$h^{-1}{\rm{M_{\odot}}}$}\fi}}
\newcommand{\Msun}{{\ifmmode{{\rm {M_{\odot}}}}\else{${\rm{M_{\odot}}}$}\fi}}

\title[Redshift remapping in dark-matter-dominated cosmological models]{Redshift remapping and cosmic acceleration in dark-matter-dominated cosmological models}

\author[R. Wojtak \& F. Prada]{Rados{\l}aw Wojtak$^{1, 2}$\thanks{E-mail: wojtak@stanford.edu, f.prada@csic.es} and Francisco Prada$^{1,3,4,5}$ \\
$^1$Kavli Institute for Particle Astrophysics and Cosmology, Stanford University, SLAC National Accelerator Laboratory, \\ Menlo Park, CA 94025, USA \\
$^2$ Dark Cosmology Centre, Niels Bohr Institute, University of Copenhagen, 
2100 Copenhagen \O, Denmark \\
$^3$ Instituto de F{\'i}sica Te{\'o}rica, (UAM/CSIC), Universidad Aut{\'o}noma de Madrid, Cantoblanco, E-28049 Madrid, Spain \\
$^4$ Campus of International Excellence UAM+CSIC, Cantoblanco, E-28049 Madrid, Spain \\
$^5$ Instituto de Astrof{\'i}sica de Andaluc{\'i}a, (IAA/CSIC), Glorieta de la Astrono{\'i}a, E-18190 Granada, Spain
\
}

\begin{document}

\maketitle

\begin{abstract}
The standard relation between the cosmological redshift and cosmic scale factor underlies cosmological inference from 
virtually all kinds of cosmological observations, leading to the emergence 
of the $\Lambda$ cold-dark-matter ($\Lambda$CDM) cosmological model. This relation is not a fundamental theory and 
thus observational determination of this function (redshift remapping) should be regarded as an insightful alternative to holding its standard 
form in analyses of cosmological data. Here we 
present non-parametric reconstructions of redshift remapping in dark-matter-dominated models and constraints on 
cosmological parameters from a joint analysis of all primary cosmological probes including the local measurement of the Hubble 
constant, Type Ia supernovae, baryonic acoustic oscillations (BAO), Planck observations of 
the cosmic microwave background (CMB) radiation (temperature power spectrum) and cosmic chronometers. 
The reconstructed redshift remapping points to an additional boost 
of redshift operating in late epoch of cosmic evolution, but 
affecting both low-redshift observations and the CMB. 
The model predicts a significant difference between the actual Hubble constant, $h=0.48\pm{0.02}$, and its local determination, 
$h_{\rm obs}=0.73\pm{0.02}$. The ratio of these two values coincides closely with the maximum expansion rate inside voids formed in the corresponding open cosmological model with $\Omega_{\rm m}=0.87\pm{0.03}$, whereas the actual value of the Hubble constant 
implies the age of the Universe that is compatible with the Planck $\Lambda$CDM cosmology. The model with redshift remapping 
provides excellent fits to all data and eliminates recently reported tensions between the Planck 
$\Lambda$CDM cosmology, the local determination of the Hubble constant and the BAO measurements from the Ly-$\alpha$ forest 
of high-redshift quasars.
\end{abstract}


\section{Introduction}

Compelling observational evidences demonstrate that the Universe is undergoing a phase of accelerated expansion. 
The accelerated expansion manifests itself in various observations probing different aspects of cosmic evolution, 
from observations of Type Ia supernovae \citep{Per1999,Rie1998}, through observations of the cosmic microwave background (CMB) radiation \citep{Pla2015}
and the abundance of galaxy clusters \citep{Man2015}. These vast observational data laid the foundation of the concordance cosmological 
model in which all observational signatures of cosmic acceleration are reconciled with the theoretical predictions 
of general relativity by introducing a positive cosmological constant $\Lambda$. The great success of the resulting 
standard $\Lambda$ cold dark matter ($\Lambda$CDM) model can be appreciated by realizing that the model is 
capable of reducing information contained in all cosmological observations to a set of only six parameters including 
the cosmological constant.

Despite its simplicity and capacity to describe key cosmological data, the  $\Lambda$CDM model lacks fundamental understanding 
of its main constituents. In particular, the cosmological constant making up nearly $70$~\% of the total energy budget in 
the Universe \citep{Pla2015} cannot be explained on the ground of well-established physics. 
It has been long known that its only viable interpretation as the vacuum energy results in the largest discrepancy between 
observations and theory which predicts its energy density exceeding the observational determinations by $\sim120$ orders 
of magnitude \citep{Car2001}.

After two decades of intense data-driven progress in cosmology, the physical nature 
of cosmic acceleration remains unknown 
and becomes arguably the most challenging mystery of modern cosmology. Many ongoing and upcoming cosmological surveys such as the Dark Energy Survey (DES) \citep{des2005}, Euclid \citep{euc2011}, the Dark Energy Spectroscopic Instrument (DESI) \citep{desi2013} or 
the Large Synoptic Survey Telescope (LSST) \citep{lsst2012} are devised 
to shed light on this problem. This observational strategy, however, does not seem to be fully balanced by concept-driven paths of 
research. In particular, the main effort focused on observational determination of dark energy equation state assumes implicitly the existence of dark energy and excludes the possibility that cosmic acceleration may be an emergent phenomenon of yet unknown physical nature. The best example of the latter possibility is a modification of gravity such as $f(R)$ gravity \citep{Clif2015}. This 
approach accounts for cosmic acceleration by modifying the Einstein's equations rather than by postulating the existence of an additional form of energy. This scenario is seriously considered as a viable alternative to dark energy, although most models of modified 
gravity must closely mimic the cosmological constant in order not to violate stringent tests of gravity on scales of the 
Solar System \citep{Bra2008,Cer2016}. Yet, it does not exhaust all possibilities for exploring the problem of cosmic acceleration as an emergent phenomenon.

Necessity of considering models going beyond the boundaries of the standard $\Lambda$CDM cosmology stems from gradually 
increasing observational anomalies. Arguably two most alarming tensions within the standard cosmological model include discrepancy 
between the local and the CMB-based Planck determination of the Hubble constant \citep{Ber2016}, and anomalous values of the baryonic acoustic oscillations (BAO) 
determined from the Ly-$\alpha$ forest of high-redshift quasars observed in the 
SDSS-III/BOSS (Sloan Digital Sky Survey/Baryon Oscillations Spectroscopic Survey) \citep{Fon2014,Del2015}. 
These two tensions are recognized at a moderate-strong confidence level of $\sim3\sigma$. However, if they are corroborated by future observations the standard model will 
need to undergo a substantial change in order to absorb both anomalies. The tension in the measurements of the Hubble constant would require a phantom-like dark energy \citep{val2016}, whereas the BAO anomaly could be solved by by employing 
models with evolving dark energy \citep{Sah2014}. Such modifications would diminish arguably appealing simplicity of 
the $\Lambda$CDM model and reinforce the motivation for exploring other theoretical propositions leading 
potentially to understanding of the cosmic acceleration phenomenon.

The property of cosmic acceleration is a necessary element in reconciling observations with the assumed theoretical framework. 
Different layers of this framework, however, can effectively play the same role in the process of establishing a concordance model. Considering two contrasting examples, cosmic acceleration can be ascribed either to dark energy or a new form of gravity. 
Less obvious layer of the framework involves the computation of cosmological observables. On one hand, this layer is expected 
to be fully specified by the assumed model of gravity. On the other hand, it may strongly depend on the accuracy of adopted 
approximate solutions. This problem of potentially inaccurate models of cosmological observables concern the standard 
cosmological model based on a fully isotropic and homogenous metric, the Friedmann-Lema$\hat{\imath}$tre-Robertson-Walker (FLRW) metric, for which potentially large differences between its derivatives and the corresponding derivatives of the actual metric 
of the Universe may result in significant discrepancies between observations and their models \citep{Green2014}. 
With the recent advent of fully relativistic cosmological simulations \citep{Ben2016,Gib2016}, we just begin 
to address this long-standing problem in a fully rigorous way.

\begin{figure*}
\centering
\includegraphics[width=0.98\textwidth]{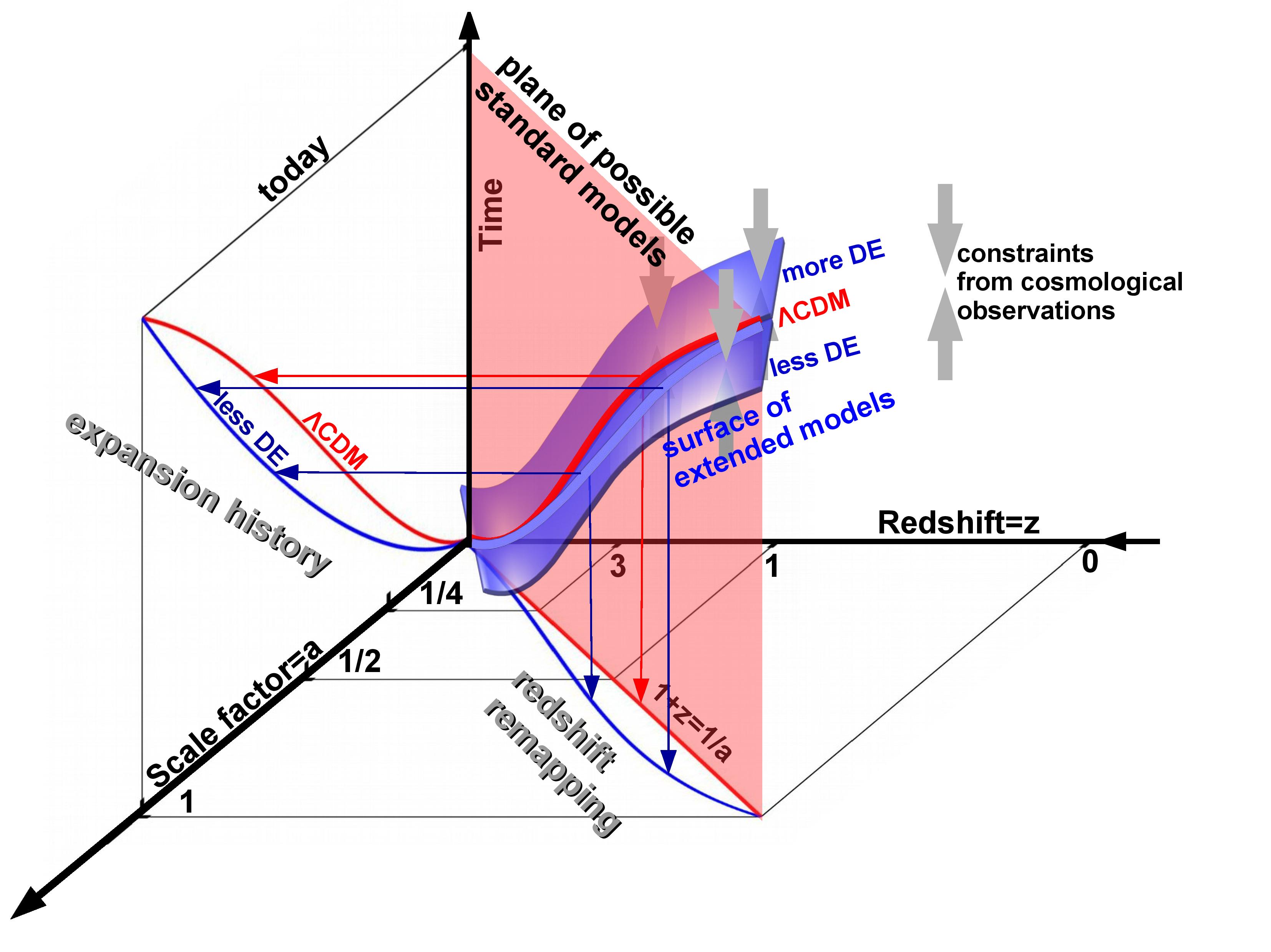} 
\caption{Conceptual scheme of cosmological models with redshift remapping. The standard approach assumes that 
mapping between the cosmological redshift $z$ and the scale factor $a$ is given by $1+z=1/a$. Therefore, standard theoretical 
models are always confined strictly to the red plane in 3D space spanned by redshift, cosmic scale factor and time. 
Redshift remapping (non-standard relation between cosmic scale factor and the cosmological redshift) violates 
this restriction and the resulting models populate space around the red plane. Cosmological 
observations constrain, however, the allowed space to a 2D manifold (the blue surface between the grey arrows 
representing constraints from cosmological data). Every model is represented by a line whose projections onto 
the $a-t$ and $a-z$ planes show respectively the expansion history and redshift remapping of the underlying model. 
Models in front of the red plane 
require less dark energy than the standard $\Lambda$CDM and $1+z>1/a$ (see an example 
shown by the blue line), whereas models behind the red plane 
are characterized by higher dark energy content and $1+z<1/a$. The intersection of the blue surface with the red plane is a 
$\Lambda$CDM model with $1+z=1/a$.
}
\label{remapping-gen}
\end{figure*}

The primary effect of cosmic evolution on light propagation is the cosmological redshift. In the standard theoretical framework 
underlying essentially all interpretations of cosmological data, the cosmological redshift is a theoretically assumed function 
of the scale factor, derived from the FLRW metric. This function, however, may not necessarily comply with the actual 
mapping between the cosmological redshift and cosmic scale factor. As shown by \citet{Bas2013} and \citet{Woj2016}, 
one can therefore attempt to reconstruct this mapping from observations rather than to postulate its theoretical form. 
In order to differentiate it from the standard relation between the cosmological redshift $z$ 
and cosmic scale factor $a$, we shall hereafter refer to observationally reconstructed $a-z$ relation as redshift remapping. 
Fig.~\ref{remapping-gen} demonstrates a geometric representation of this idea in space spanned by 
cosmic time, cosmic scale factor and the cosmological redshift. The red plane contains all theoretical models with the 
standard mapping between cosmic scale factor and the cosmological 
redshift $(1+z=1/a)$, i.e. with the same projection onto the $a-z$ plane. 
Cosmological observations reduce possible solutions to a line whose projection onto the $a-t$ plane is the expansion 
history of the standard $\Lambda$CDM model. Including redshift remapping as an additional degree of freedom opens up 
new space of models. As shown in our recent work \citep{Woj2016}, observational constraints on redshift remapping 
from low-redshift probes, including Type Ia supernovae and BAO, are degenerated with the dark energy density parameter.  
Therefore, the new cosmological models with redshift remapping form a 2D manifold 
of models (the blue surface in Fig.~\ref{remapping-gen}) 
with different combinations of dark energy (expansion history) and redshift remapping. In the context 
of cosmic acceleration, of particular interest is the model dominated by dark matter (without dark energy) 
in which the observationally determined redshift remapping accounts for cosmic 
acceleration. In the above description, the 2D manifold is an idealized representation of the posterior 
probability distribution which would effectively gives rise to a small thickness of this surface. Without loss 
of generality, this idealization is made for the sake of clarity of demonstrating new cosmological inference 
with redshift remapping.

The current cosmological models with redshift remapping are phenomenological in a sense that 
the $a-z$ relation is constrained by observations amongst standard cosmological parameters. However, there clearly exists 
a potential for theoretical considerations. Non-trivial redshift remapping can emerge both in non-metric theories of 
gravity such as $\tilde{\delta}$ gravity \citep{Alf2012,Alf2013} and many general-relativistic models 
going beyond the standard FLRW cosmology, e.g. the Lindquist-Wheeler model with a discretized matter content 
\citep{Cli2009,Cli2012}, conformal FLRW cosmologies \citep{Vis2015}, models with non-linear inhomogeneities 
\citep[see e.g.][]{Meu2012, Lav2013}. The latter group of models is probably much broader. We enumerate 
only selected examples for which a deviation of the $a-z$ relation from its standard form was explicitly shown.

The main goal of this paper is to improve the results presented in \citep{Woj2016} by combining all relevant low-redshift 
probes and CMB observations in a joint analysis of models with redshift remapping. Including CMB data not only 
improves observational constrains, but it also allows for revisiting anomalies in the Hubble constant determinations 
and the BAO observations in the context of new cosmological models with redshift remapping. In addition, we develop a new 
method which allows for a non-parametric reconstruction of redshift remapping from observations. The 
new method replaces a simplified parametric approach employed in \citep{Woj2016}. We limit our analysis 
to the most interesting case of dark-matter-dominated models without dark energy. Therefore, the final 
constraints on cosmological parameters and redshift remapping establish an alternative model in which 
cosmic acceleration emerges exclusively from a new phenomenological mapping between the cosmological 
redshift and cosmic scale factor.

The manuscript is organized as follows. In section 1, we introduce the idea of redshift remapping and describe a 
new approach to its observational determination based on a non-parametric fitting. In section 2, we outline the 
framework for computing all relevant observables in cosmological models with redshift remapping. Next, we 
describe cosmological data used in this work. In section 3, we present observational constraints on cosmological 
parameters and redshift remapping. Then we discuss all results in section 4 and summarize in section 5.

\section{Redshift remapping}
Extragalactic redshifts in the CMB rest frame $z_{\rm obs}$ can be factorized into the contributions from expansion of space 
along the photon path $z_{\rm s}$ and the peculiar velocity of the emitter in the following way
\begin{equation}
1+z_{\rm obs}=(1+z_{\rm s})(1+v_{\rm pec}/c),
\end{equation}
where $v_{\rm pec}$ is the projection of the peculiar velocity onto the line of sight of the observer. 
Without loss of generality, we assume here that our 
observations do not include objects with strong gravity such as black holes and we also neglect higher order relativistic corrections from 
large scale structures such as gravitational redshift (although imprints of these effects are already detectable in the current cosmological 
surveys, see \citep{Woj2011,Jim2015,Sad2015}). Peculiar velocities play important role in the interpretation of the observation of the local Universe where measured redshifts are of 
comparable magnitude to Doppler shifts due to peculiar velocities. However, in many cases of high redshift observations their contribution 
can be neglected and the observed redshift $z_{\rm obs}$ can be identified with $z_{\rm s}$. This approach is fully justified when the only 
effect of peculiar velocities is to give rise an additional scatter in multiple measurements around a relation between a cosmological 
observable and $z_{\rm s}$ (for example, the Hubble diagram for Type Ia supernovae) or when the effect vanishes as a result of averaging in redshift bins (for example, redshift bins of BAO measurements). Since all data sets used in our work fall into one of the two 
above categories, we hereafter assume $z_{\rm obs}=z_{\rm s}$, as it is customary to proceed in all standard cosmological analyses 
unless the peculiar velocity field itself is an observable.

Cosmological observations constrain in general theoretical relations between various cosmological observables 
and the cosmological redshift $z_{\rm obs}$ which reflects the expansion of space integrated along the photon path. 
Utilizing these observations to determine the expansion history and the contents of the Universe requires two 
assumptions: the mapping between $z_{\rm obs}$ and the scale factor $a$ describing global evolution of cosmic space 
and an equation relating time evolution of cosmic scale factor $a(t)$ to the energy-matter contents of the Universe. Within the standard cosmological framework, both relations are 
derived assuming a fully isotropic and homogeneous metric: the Friedmann-Lema$\hat{\imath}$tre-Robertson-Walker (FLRW) 
metric,
\begin{equation}
\textrm{d}s^{2}=\textrm{d}t^{2}-a(t)^{2}\Big(\frac{\textrm{d}r^2}{1-kr^{2}}+r^{2}\textrm{d}\Omega\Big),
\end{equation}
where $k=\pm1,\;0$. The mapping between $a$ and $z_{\rm obs}$ in this case is reduced to the well-known expression
\begin{equation}
1+z_{\rm obs}=\frac{1}{a}
\label{standard-map}
\end{equation}
and the time evolution of the scale factor is governed by the Friedmann equation, i.e.
\begin{equation}
\Big(\frac{\dot{a}}{a}\Big)^{2}/H_{0}^{2} \equiv \frac{H^{2}}{H_{0}^{2}}=\Omega_{\rm R}a^{-4}+\Omega_{\rm m}a^{-3}+\Omega_{\rm K}a^{-2}+\Omega_{\rm \Lambda},
\label{friedeq}
\end{equation}
where $H \equiv \dot{a}/a$ is the Hubble parameter and $H_{0}$ is the
Hubble constant today when $a \equiv 1$. $\Omega_{\rm R}$,
$\Omega_{\rm m}$, $\Omega_{\rm K}$, and $\Omega_{\rm \Lambda}$ are
the radiation, matter (both cold and baryonic), spatial curvature and
cosmological constant density parameters.

The key notion of our model was introduced in \citep{Bas2013,Woj2016} and is based  on the realization 
that the mapping between the scale factor $a$ and the cosmological redshift $z_{\rm obs}$ does 
not have to be a necessary assumption in cosmological inference, but instead can be observationally determined, 
simultaneously with other relevant cosmological parameters. It is quite intuitive to expect that the empirically 
motivated $a-z_{\rm obs}$ mapping should not 
deviate significantly from the standard mapping given by equation (\ref{standard-map}). It is therefore reasonable to quantify 
the new mapping as a relative correction to the standard relation based on the FLRW metric. Following \citep{Bas2013} and 
\citep{Woj2016}, we use the ratio of the redshift inferred from the FLRW metric to the actually observed redshift
\begin{equation}
\frac{z_{\rm FLRW}}{z_{\rm obs}}\equiv\frac{1/a-1}{z_{\rm obs}}=1+\alpha(z_{\rm obs}),
\label{remapping}
\end{equation}
where
\begin{equation}
1+z_{\rm FLRW}\equiv \frac{1}{a}.
\label{zFLRW-def}
\end{equation}
and $\alpha(z_{\rm obs})$ is a free function of $z_{\rm obs}$. Using the above ratio of the two redshifts allows to avoid an unphysical situation with non-vanishing $z_{\rm FRLW}$ redshifts at distances approaching $0$. For any finite value of $\alpha$ at $z_{\rm obs}=0$, $z_{\rm obs}=0$ implies vanishing $z_{\rm FLRW}$.

The function $\alpha(z_{\rm obs})$ in equation (\ref{remapping}) describes a relative difference between the standard mapping given by 
equation (\ref{standard-map}) and observationally motivated mapping between the scale factor and the observed redshift 
$z_{\rm obs}$. Hereafter we shall refer to the latter as \textit{redshift remapping}. The standard mapping and redshift remapping 
become identical when $\alpha(z_{\rm obs})=0$.

Every mapping between the scale factor and the observed redshift is always ascribed to gravity and the metric which determine 
propagation of photons. Therefore, redshift remapping can emerge either from differences between the true metric of the Universe 
and the FLRW approximation or from violation of the principle describing photon paths as null geodesic lines in spacetime. As we shall 
see in the following section, the way how we calculate all cosmological observables in our model \textit{excludes the possibility 
of interpreting redshift remapping as a non-metric-based effect} such as Zwicky's hypothesis of 'tired light`. This important 
property allows us to comply with cosmological observations ruling out non-metric-based origin of cosmological redshifts, 
e.g. redshift dependence of the spectral aging rate of Type Ia supernovae \citep{Blo2008} or blackness of the CMB spectral 
energy distribution \citep{Ell2013}.

\subsection{Parametrization}

The main difficulty of constraining redshift remapping lies in the fact that it employs an unknown function $\alpha(z_{\rm obs})$ that increases substantially the degrees of freedom of cosmological fits. As an attempt to resolve this problem \citep{Bas2013} proposed 
simple parametrizations based on Taylor expansion of $\alpha(z_{\rm obs})$ and \citep{Woj2016} used a well-motivated but 
simplistic ansatz for redshift remapping. Although both approaches can give many interesting insights into cosmological 
fits with redshift remapping, they do not provide a complete solution of the problem. In particular, they do not allow to constrain 
the shape of redshift remapping.

The best available technique which can be used to reconstruct the full shape of redshift remapping from observations 
is a non-parametric fit. Here we describe several technical details how we implement this technique in our analysis.

The current cosmological data including observations of Type Ia supernovae and BAO allow us to probe redshift remapping 
in the redshift range from $z_{\rm obs}=0$ to $z_{\rm obs}\approx2.5$. The upper limit comes from the BAO 
measurements based on observations of the Ly-$\alpha$ 
forest of high-redshift SDSS-III/BOSS quasars. A non-parametric fit relies on approximating an unknown function by a set of its values 
at several points of an independent variable and adopting an interpolation scheme for evaluating the function between these points. 
Performing a series of trial fits, we found that the most optimized set-up for a non-parametric reconstruction of redshift remapping 
employs spline interpolation between four redshifts given by $z_{\rm obs}=0,\;0.5,\;1.0,\;2.5$. The last redshift bin includes both high redshift 
supernovae, Hubble parameter from cosmic chronometers and the BAO measurements from observations of the Ly-$\alpha$ 
forest of BOSS quasars. Redshift remapping $\alpha(z_{\rm obs})$ is therefore fully determined by 6 parameters: 
4 values of $\alpha(z_{\rm obs})$ at the four redshifts $z_{\rm obs}$ used by the interpolation scheme and 2 values of derivative $\textrm{d}\alpha(z_{\rm obs})/\textrm{d}z_{\rm obs}$ at $z_{\rm obs}=0,\;2.5$. As a sanity check, we verified that our constraints on redshift remapping 
barely depend on the actual choice of binning adopted for the interpolation. In particular, changing the bin sizes or the position 
of the last bin keep the best-fitting profile of redshift remapping well within the error envelope of our main results. Moreover, increasing 
the number of bins does not reveal any additional and statistically significant features of the profile (see Fig.~\ref{alpha-profile-bins} in Appendix).

The employed technique of constraining redshift remapping allows in general for a number of solutions which can be rejected 
on the ground of expecting a relatively simple functional form of $\alpha(z_{\rm obs})$. For example, oscillatory solutions or 
highly non-monotonic functions seem to be highly suspicious and, on the basis of the argument from simplicity, they should rather 
be ignored. To comply with this condition we narrow down the whole family of solutions by means of employing prior 
distributions on the six parameters determining redshift remapping. Our choice of the priors is motivated by the expected 
form of redshift remapping for dark-matter-dominated cosmological models considered in this work. As shown in \citep[][]{Woj2016}, 
redshift remapping in this case increases monotonically with redshift and $1+\alpha(z_{\rm obs})\leq 1$. Adopting this 
property as a prior in our analysis leads to the following condition
\begin{equation}
\alpha(0)\leq\alpha(0.5)\leq\alpha(1.0)\leq\alpha(2.5) \leq 0.
  \label{prior}
\end{equation}
As we shall see in the following section, the above inequality implies automatically monotonicity 
of the comoving distance as a function of redshift $z_{\rm obs}$. However, they do not guarantee that the Hubble parameter as a function redshift is free of wiggle-like features. Therefore, as the second prior we require every model to comply with the condition 
that the Hubble parameter increases monotonically with redshift. This condition is tested numerically in every iteration 
of our cosmological fits. Due to the relatively weak constraining power of the data in the last redshift bin, 
we also narrow down the range of the first derivative of $\alpha(z_{\rm obs})$ at $z_{\rm obs}=2.5$ to $[0,0.1]$.

The CMB observations provide additional constraints on $\alpha(z_{\rm obs})$ at the redshift of the last scattering surface, 
$z_{\rm obs}\approx 1100$. This measurement probes a single 
value of redshift remapping and it is independent of the constraints from low 
redshift probes (apart from the case of a joint analysis combining low-redshift and CMB data). 
Comparing constraints on $\alpha(z_{\rm obs})$ from 
low-redshift probes with its value at the last scattering surface can give some hints on the form of 
$\alpha(z_{\rm obs})$ in the redshift range without any data. For example, if the former indicates a clear convergence 
of $\alpha(z_{\rm obs})$ and its maximum value at $z_{\rm obs}<2.5$ is consistent with redshift remapping determined 
from the CMB then one can likely expect the existence of a plateau extending between $z_{\rm obs}=2.5$ and 
$z_{\rm obs}\approx 1100$.

\subsection{Background model}
In order to close the system of equations underlying the framework for calculating all basic cosmological observables 
such as distances, we need to specify the equation governing the time evolution of the scale factor. Despite the fact 
that our approach modifies the standard mapping between the scale factor and the observed redshift -- one of the direct 
predictions based on the FLRW metric, we keep the Friedmann equation given by (\ref{friedeq}) unchanged. Since 
we restrict our work to dark-matter-dominated cosmological models ($\Omega_{\rm \Lambda} \equiv 0$), the Friedman equation at redshifts characteristic 
for BAO and Type Ia supernova observations reads
\begin{equation}
\frac{(\dot{a}/a)^{2}}{H_{0}^{2}}\equiv \frac{H^{2}}{H_{0}^{2}}=\Omega_{\rm m}(1+z_{\rm FLRW})^{3}+(1-\Omega_{\rm m})(1+z_{\rm FLRW})^{2}.
\label{friedmann}
\end{equation}

Combining the Friedmann equation with a modification of the standard redshift remapping may not seem to form a self-consistent 
framework. This inconsistency is inevitable when we assume that the FLRW metric is accurate on all scales and general relativity 
is correct at least on large scales. However, this condition can be broken either due to inadequacy of the FLRW metric for 
describing the first derivatives of the actual metric of the Universe \citep{Green2014} or due to the violation of a general framework 
of relativity in which photons propagate along null geodesic lines. The former circumstance arises in many general-relativistic cosmological 
models whose global evolution is governed by the Friedmann equation, but whose observables cannot be accurately described 
with the FLRW metric \citep[see e.g.][]{Cli2009,Lav2013,Ben2016}. 
The key optical property undergoing substantial modification compared to the predictions based on the FLRW metric 
is the mapping between the scale factor and the cosmological redshift.

Redshift remapping is a phenomenological model which does not attempt to incorporate any specific theoretical model predicting 
deviations from the standard mapping between the scale factor and the cosmological redshift based on the FLRW metric. 
Keeping the phenomenological status of our model allows us to be independent of any specific theoretical 
explanations of redshift remapping, no matter if they are related to the inadequacy of the FLRW metric or some modifications 
of general-relativistic principles of light propagation. However, due to its simplicity and the potential for a straightforward and systematic exploration, we tend to think of redshift remapping as a phenomenon emerging from differences between the true metric of the Universe 
and the FLRW approximation.

With the recent advent of fully relativistic cosmological simulations based on obtaining exact numerical solutions of the Einstein field equations \citep{Ben2016,Gib2016}, effects of structure formation on the actual metric can be studied in a much more rigorous way than before. Recent studies of such simulations demonstrated that non-linear evolution of cosmic structures induces 
strong inhomogeneities in the local expansion rate what is expected to have a non-negligible effect of the computation of 
observables, in particular the mapping between the cosmological redshift and cosmic scale factor. At the same time, the presence of 
such a locally inhomogeneous expansion appears to have a negligible impact on the global expansion of the Universe. 
Combining these two facts we can see that our new framework based on the Friedmann equation and redshift remapping 
can be fully consistent with general relativity. In this interpretation, redshift remapping is merely a theoretically unknown 
relation between the cosmological redshift and cosmic scale factor, which has yet to be established in upcoming advances 
of fully relativistic simulations.

\section{Observables}

We calculate all cosmological observables assuming that redshift remapping is a purely metric-based effect. This assumption implies 
that the distance-duality relation with the observed redshift $z_{\rm obs}$ holds, even though $z_{\rm obs}$ can be different than 
$z_{\rm FLRW}$ \citep{Meu2012},
\begin{equation}
D_{\rm L}(z_{\rm obs})=(1+z_{\rm obs})^{2}D_{\rm A}(z_{\rm obs}),
\label{distance-duality}
\end{equation}
where $D_{\rm L}$ is the luminosity distance and $D_{\rm A}$ is the angular diameter distance. The relation follows directly 
from two fundamental principles: the conservation of photons and the so-called reciprocity theory (also known 
as Etherington's theory, \citep{Eth1933}). Needless to say, the relation would be violated if $(1+z_{\rm obs})$ 
factor was replaced by $(1+z_{\rm FLRW})$.

\subsection{Cosmological distances}
We assume that the comoving distance $D_{\rm M}$ to an object observed at redshift $z_{\rm obs}$ is given by
\begin{equation}
D_{\rm M}(z_{\rm obs})=c\int_{0}^{z_{\rm obs}[1+\alpha(z_{\rm obs})]}\frac{\textrm{d}z_{\rm FLRW}}{H(z_{\rm FLRW})},
\label{d-com}
\end{equation}
where $H(x)$ is the Hubble parameter defined in equation (\ref{friedmann}). Compared to the standard formula based 
on the FLRW metric (see e.g. \citep{Hog1999}), the only difference arising from redshift remapping concerns 
the upper limit of the integral which relates the observed redshift to the scale factor at the time of emission. 
The standard formula is recovered when $\alpha(z_{\rm obs})$ vanishes at all redshifts.

As a consequence of the distance-duality relation, redshift remapping does not modify the standard way of deriving the 
luminosity and angular diameter distances and they both are calculated as follows,
\begin{eqnarray}
D_{\rm L}(z_{\rm obs})  & = &  (1+z_{\rm obs}) \, D_{\rm H} \, f_{C}[D_{\rm M}(z_{\rm obs})/D_{\rm H} ]\nonumber \\
D_{\rm A}(z_{\rm obs})  & = &  \frac{1}{1+z_{\rm obs}} \, D_{\rm H} \, f_{C}[D_{\rm M}(z_{\rm obs})/D_{\rm H}],
\label{dLdA}
\end{eqnarray}
where $D_{\rm H}=(c/H_{0})/\sqrt{|\Omega_{k}|}$ and
\begin{equation}\label{f(x)}
	f_{C}(x)=  \left\{
	\begin{array}{lll}
	 \sinh(x) & \Omega_{\rm k}>0,\\
	 x & \Omega_{\rm k}=0,\\
	 \sin(x) & \Omega_{\rm k}<0.\\
\end{array} \right.
\end{equation}
It is easy to check that both distances conform to the postulated distance-duality relation given by \ref{distance-duality}. 
$D_{\rm L}$ and $D_{\rm A}$ are measured from Type Ia supernova and BAO data, respectively.

\subsection{Hubble parameter}

The local determination of the Hubble constant and the measurements of the Hubble parameter from BAO observations 
rely in general on the same key idea of measuring the ratio of the redshift (velocity) intervals to the corresponding 
distance scales. Measurements of distances make use of various external distance calibrators such as cepheids 
(for the local measurements) or the BAO scale from the CMB (for the BAO-based measurements). Therefore, they are 
obviously independent of redshift remapping. It is then easy to notice that redshift remapping modifies 
the measured Hubble parameter only through the observed redshift interval $\textrm{d}z_{\rm obs}$, i.e. 
$H_{\rm obs}(z_{\rm obs})\sim \textrm{d}z_{\rm obs}$.

In the framework built upon redshift remapping, the observationally determined Hubble parameter is no longer identical 
to the Hubble parameter in the Friedmann equation. Hereafter, we shall refer the former as 
$H_{\rm obs}(z_{\rm obs})$ and to the latter as $H(z_{\rm FLRW})$. The relation between both of them 
can be easily found by transforming the observed redshift interval $\textrm{d}z_{\rm obs}$ into the corresponding 
interval $\textrm{d}z_{\rm FLRW}$:
\begin{equation}
H_{\rm obs}(z_{\rm obs})=\frac{H[z_{\rm FLRW}(z_{\rm obs})]}{\textrm{d}z_{\rm FLRW}/\textrm{d}z_{\rm obs}} 
= \frac{H[z_{\rm obs}(1+\alpha)]}{1+\alpha+z_{\rm obs}\textrm{d}\alpha/\textrm{d}z_{\rm obs}}.
\label{Hub}
\end{equation}
As a special case of the above equation, we can see that redshift remapping implies that the locally measured Hubble constant,
\begin{equation}
H_{\rm obs\;0}=H_{\rm obs}(z_{\rm obs}=0)=h_{\rm obs}100\;{\rm km~s}^{-1}{\rm ~Mpc}^{-1},
\end{equation}
cannot be identified with the actual Hubble constant,
\begin{equation}
H_{\rm 0}=H(z_{\rm obs}=0)=h100\;{\rm km~s}^{-1}{\rm ~Mpc}^{-1},
\end{equation}
related to the global expansion of space. Using 
$z_{\rm obs}=0$ we find that
\begin{equation}
H_{\rm obs\;0}=\frac{H_{0}}{1+\alpha(0)}.
\label{hub-loc-obs}
\end{equation}
With the adopted priors on redshift remapping motivated by cosmological fits with dark-matter-dominated models, we 
expect $H_{\rm obs\;0}\leq H_{0}$.

Some measurements of the BAO signal constrain the volume averaged distance combining the observed 
Hubble parameter and the angular diameter distance. This distance is given by the following formula
\begin{equation}
D_{\rm V}(z_{\rm obs})=\Big(\frac{D_{\rm A}^{2}(z_{\rm obs})}{cH_{\rm obs}(z_{\rm obs})}\Big)^{1/3}.
\end{equation}
Dependence on redshift remapping occurs via equations (\ref{dLdA}) and (\ref{Hub}).

The Hubble parameter can be estimated from observations of cosmic chronometers which are pairs of passively evolving galaxies separated in redshift space by a small range ${\rm d} z_{\rm obs}$ \citep{Jim2002}. 
Passive evolution of both galaxies allows for estimating the corresponding cosmic time interval ${\rm d}t$ as a difference between 
stellar ages of the two galaxies. The resulting ratio of the redshift-to-time interval constrains the expansion rate via the following 
equation:
\begin{equation}
H_{\rm CC}(z_{\rm obs})=-\frac{1}{1+z_{\rm obs}}\frac{{\rm d}z_{\rm obs}}{{\rm d}t}.
\end{equation}
The above observable equals to the Hubble parameter $H(z_{\rm obs})$ given by the Friedmann equation when the standard 
mapping between $z_{\rm obs}$ and the scale factor is assumed, i.e. ($z_{\rm FLRW}=z_{\rm obs}$). Redshift remapping, 
however, leads to the following modification:
\begin{equation}
H_{\rm CC}(z_{\rm obs})=\frac{H[z_{\rm FLRW}(z_{\rm obs})]}{1+z_{\rm obs}}
\frac{1+z_{\rm FLRW}}{1+\alpha+z_{\rm obs}{\rm d}\alpha/{\rm d}z_{\rm obs}}.
\end{equation}

\subsection{Cosmic Microwave Background}

We begin our considerations with a general statement that redshift evolution of the CMB 
temperature can be derived without any assumption about the mapping between 
the cosmological redshift and cosmic scale factor, and thus it does not depend on redshift remapping. In fact, the standard 
relation between the CMB temperature and the observed redshift, i.e. $T\sim (1+z_{\rm obs})$, 
and the conservation of the blackbody spectrum are equivalent to the distance-duality relation \citep{Ras2016}. Since 
the distance-duality relation holds in our model, the standard dependence of the CMB temperature 
on $z_{\rm obs}$ as well as the blackness of the CMB spectrum, the former confirmed by observations 
of the Sunyaev-Zel'dovich effect \citep{Sar2014,Luz2015} and the latter tested in the measurements of the 
spectral energy distribution of the CMB \citep{Ell2013,Fix1996}, remain unchanged, no matter what 
mapping between $z_{\rm obs}$ and cosmic scale factor $a$ is adopted.

The present temperature of the CMB fixes the redshift of recombination at $z_{\rm obs\;rec}\approx 1100$. 
Finding the scale factor at the epoch of recombination requires going beyond the observational basis and 
assuming a mapping between $z_{\rm obs}$ and the corresponding scale factor. In contrast to the standard 
cosmological framework in which $a=1/(1+z_{\rm obs})$, this scale factor is effectively a free parameter 
in our model (an unconstrained value of redshift remapping at the redshift of recombination). Therefore, 
the main effect of redshift remapping on the CMB modeling is to change the scale factor of recombination 
from $a_{\rm rec}=1/(1+z_{\rm obs\;rec})$ to 
\begin{equation}
a_{\rm rec}\equiv \frac{1}{1+z_{\rm FLRW\;rec}}=\frac{1}{1+z_{\rm obs\;rec}[1+\alpha(z_{\rm obs\;rec})]}.
\end{equation}

We use the \textit{camb} code \citep{Lew1999} to compute the CMB temperature (TT) power spectrum. The 
standard mapping between the 
scale factor and the cosmological redshift is a built-in relation in the code which is used to determine the scale factor of 
recombination given the measured temperature of the CMB $T_{\rm obs\;0}$. An easy way to incorporate 
redshift remapping into the framework of the code is to replace the input CMB temperature in \textit{camb} 
by an auxiliary temperature $T_{\rm 0}$ given by
\begin{equation}
T_{\rm 0}=T_{\rm obs\;0}\frac{1+z_{\rm obs\;rec}}{1+z_{\rm FLRW\;rec}}\approx T_{\rm obs\;0}\frac{1}{1+\alpha(z_{\rm obs\;rec})}.
\end{equation}
The auxiliary temperature is effectively a free parameter which sets the scale factor of recombination based on the standard $a-z_{\rm rec}$ mapping, i.e.
\begin{equation}
a_{\rm rec}=\frac{T_{\rm 0}}{T_{\rm rec}},
\end{equation}
where $T_{\rm rec}$ is the CMB temperature at the epoch of recombination. The power spectrum calculated 
for the auxiliary temperature does not describe the actual measurements of the CMB temperature anisotropies. 
It merely represents the power spectrum of a model with the standard redshift mapping and the current CMB 
temperature equal to the auxiliary temperature $T_{\rm 0}$. In order to make it represent the actual CMB 
observations, it has to be transformed according to the conversion of the CMB temperature from the 
auxiliary temperature $T_{\rm 0}$ back to the observed temperature $T_{\rm obs\;0}$. This transformation 
involves both scaling the CMB temperature fluctuation $\Delta T$ and the angular diameter distance 
$D_{\rm A}$ to the last scattering surface in the following way:
\begin{eqnarray}
\Delta T & \rightarrow & \Delta T(T_{\rm obs\;0}/T_{\rm 0}) \nonumber \\
D_{\rm A} & \rightarrow & D_{\rm A}\frac{1+z_{\rm FLRW\;rec}}{1+z_{\rm obs\;rec}}=D_{\rm A}(T_{\rm obs\;0}/T_{\rm 0}). \nonumber \\
\label{conv1}
\end{eqnarray}
The latter scaling stems directly from the distance-duality relation in which the conversion from the auxiliary 
to the actual CMB temperature results in replacing the factor $(1+z_{\rm FLRW})$ by 
$(1+z_{\rm obs})=(1+z_{\rm FLRW})(T_{\rm 0}/T_{\rm obs\;0})$. Using the above conversions of angular diameter 
distance and the CMB temperature we find that the power spectrum $C_{l}$ and the multipole numbers 
transform in the following way:
\begin{eqnarray}
C_{l} & \rightarrow & C_{l}(T_{\rm obs\;0}/T_{\rm 0})^{2} \nonumber \\
l & \rightarrow & l(T_{\rm obs\;0}/T_{\rm 0}) \nonumber \\
\label{conv2}
\end{eqnarray}
The whole transformation is independent of the order of these two scaling relations given above. The transformation of the 
multipole numbers assumes the approximation of small angles. Therefore, we limit our analysis of the CMB temperature 
power spectrum to high multipoles with $l\geq30$.

The procedure outlined above becomes very intuitive in a plausible scenario in which redshift remapping 
is related to a physical effect operating at late times of cosmic evolution, i.e. the epoch of non-linear 
structure formation. In this case, a non-vanishing $\alpha(z_{\rm obs\;rec})$ is simply a residual effect of the propagation of the 
CMB photons at late times. Thus, this effect does not coincide with the physics of the early universe determining the CMB 
properties and it can be separated from the model by applying well-defined transformations of the CMB temperature 
and the angular scales of the CMB temperature anisotropies. It is not guaranteed, however, that cosmological fits with redshift 
remapping do not require some modifications of the CMB lensing. In order to keep the model as general as possible, 
we treat the lensing amplitude in \textit{camb} as a free parameter (see below).

\section{Data sets and methods}
We distinguish two main groups of cosmological data sets: low-redshift probes (LZ) and high-redshifts probe (HZ). The former 
constrains redshift remapping at redshifts $z_{\rm obs}$ between 0 and $2.5$ and includes observations of Type Ia supernovae, 
BAO, cosmic chronometers and the local measurement of the Hubble constant. The latter places constraints on redshift remapping 
at the redshift of recombination and includes the power spectrum of the CMB temperature anisotropies.

\begin{table*}
\begin{center}
\begin{tabular}{ccccc}
$z_{\rm obs}$ & observable & measurement & survey & reference \\
\hline
 & & &  & \\
$0.106$ & $D_{\rm V}$ [Mpc] & $439$ & 6dF & \citep{Beu2011} \\
$0.15$ & $D_{\rm V}$ [Mpc] & $660$ & MGS & \citep{Ros2015} \\
$0.38$ & $D_{\rm M}$ [Mpc] & $1529$ & BOSS & \citep{Ala2016} \\
$0.38$ & $H_{\rm obs}$ [km/s/Mpc] & $81.2$ & BOSS & \citep{Ala2016} \\
$0.44$ & $D_{\rm V}$ [Mpc] & $1707$ & WiggleZ & \citep{Kaz2014} \\
$0.51$ & $D_{\rm M}$ [Mpc] & $2007$ & BOSS & \citep{Ala2016} \\
$0.51$ & $H_{\rm obs}$ [km/s/Mpc] & $88$ & BOSS & \citep{Ala2016} \\
$0.60$ & $D_{\rm V}$ [Mpc] & $2209$ & WiggleZ & \citep{Kaz2014} \\
$0.61$ & $D_{\rm M}$ [Mpc] & $2274$ & BOSS & \citep{Ala2016} \\
$0.61$ & $H_{\rm obs}$ [km/s/Mpc] & $96$ & BOSS & \citep{Ala2016} \\
$0.73$ & $D_{\rm V}$ [Mpc] & $2502$ & WiggleZ & \citep{Kaz2014} \\
$2.34$ & $D_{\rm A}$ [Mpc] & $1666$ & BOSS(Ly-$\alpha$) & \citep{Del2015} \\
$2.34$ & $H_{\rm obs}$ [km/s/Mpc] & $221$ & BOSS(Ly-$\alpha$) & \citep{Del2015} \\
$2.36$ & $D_{\rm A}$ [Mpc] & $1593$ & BOSS(Ly-$\alpha$) & \citep{Fon2014} \\
$2.36$ & $H_{\rm obs}$ [km/s/Mpc] & $225$ & BOSS(Ly-$\alpha$) & \citep{Fon2014} \\
  & & &  & \\  
\end{tabular}
\caption{BAO measurements (best-fit values) used in this work. All results are calibrated with the 
same fiducial value of the sound horizon $r_{\rm s}=147.78$~Mpc.}
\label{bao-table}
\end{center}
\end{table*}

In the LZ data set, we use apparent magnitudes of Type Ia supernovae resulting from a joint likelihood analysis (JLA) 
of SDSS-II and SNLS supernova samples \citep{Bet2014}. The data set consists of supernova magnitudes averaged in 31 redshift bins and the associated 
covariance matrix. Second, we use a compilation of BAO measurements combining results from fours spectroscopic 
surveys: SDSS-III/BOSS \citep{Ala2015}, WiggleZ \citep{Par2012}, SDSS-MGS \citep{Aba2009} and 6dF \citep{Jon2009}. 
The BOSS data set includes constraints on the Hubble parameter and angular 
diameter distance inferred from galaxy clustering in three redshift bins centered at $z_{\rm obs}=0.38,0.51,0.61$, 
flux-correlation of the Ly-$\alpha$ forest of BOSS quasars at effective redshift $z_{\rm obs}=2.34$ \citep{Del2015} 
and from the cross-correlation of quasars with the Ly-$\alpha$ forest at effective redshift $z_{\rm obs}=2.36$ \citep{Fon2014}. 
For the three low redshift bins, we use the consensus best-fit 
values and the covariance matrix obtained by combining results of four independent pre-reconstruction 
methods of measuring the BAO signal \citep{Ala2016}. The two BAO measurements at high redshifts are 
practically uncorrelated \citep{Del2015} and therefore both are described by two independent covariance matrices. 
Since $H_{\rm obs}$ is only mildly anticorrelated with $D_{\rm A}$ in both cases, we use diagonal covariance 
matrices. The WiggleZ data provides the measurements of the volume-averaged distance $D_{\rm V}$ in three redshift 
bins at effective redshifts $z_{\rm obs}=0.44,0.60, 0.73$ \citep{Kaz2014}. Our analysis takes into account the full 
covariance matrix quantifying correlations arising from partial overlaps between redshift bins. Finally, 
we also make use of single measurements of $D_{\rm V}$ from the 6dF survey at effective redshift 
$z_{\rm obs}=0.106$ \citep{Beu2011} and from the Main Galaxy Sample of the SDSS (SDSS-MGS) at effective redshift 
$z_{\rm obs}=0.15$ \citep{Ros2015}. All BAO measurements used in this work are summarized in Table~\ref{bao-table} 
and shown in Fig. \ref{res-LZ}. They are consistently rescaled to the same value of the sound horizon 
$r_{\rm s\;fid}=147.78$~Mpc assumed in \citep{Ala2016} for a fiducial flat $\Lambda$CDM cosmological model 
with $\Omega_{\rm m}=0.31$, baryon density $\Omega_{\rm b}h^{2}=0.022$, Hubble constant $h=0.676$, 
fluctuation amplitude $\sigma_{8}=0.83$, spectral tilt $n_{\rm s}=0.97$ and the reionization optical depth 
$\tau=0.078$. The fiducial cosmological parameters are within $1\sigma$ range of the best-fit Planck2015 
values \citep{Pla2015}. Without loss of generality, we hereafter refer to the fiducial model as the Planck cosmological 
model.

The LZ data set is complemented by the local measurement of the Hubble constant from \citep{Rie2016}, i.e. 
$H_{\rm obs\;0}=(73.24\pm1.74)$~km~s$^{-1}$~Mpc$^{-1}$. Including this measurement in our analysis 
allows us to place constraints on $r_{\rm s}$ independently of CMB observations. Finally, we also include 
the estimates of the Hubble parameter $H_{\rm CC}$ from observations of cosmic chronometers (see 
Table \ref{hubpar-cc}).

The HZ data set includes the temperature power spectrum from the Planck 2015 data release \citep{Pla2015a}. 
We compute the likelihood function using the standard code provided as a part of the Planck Legacy Archive. 
Due to small-angle approximation assumed in equation~(\ref{conv2}), we restrict our analysis to high multipoles 
with $l\geq30$ available in a data set named \textit{plik\textunderscore{}dx11dr2\textunderscore{}HM\textunderscore{}v18\textunderscore{}TT.clik}. The likelihood function depends on a theoretical power spectrum and $16$ nuisance parameters 
describing the foreground sources. The use the recommended priors for the foreground parameters.

We use a Monte Carlo Markov Chain technique to find best-fitting models. The likelihood function for the LZ dat set 
is given by $\exp(-\chi^{2}/2)$, where $\chi^{2}$ function is computed using the covariance matrix whose blocks 
are given by covariance matrices of the subsequent data subsets.

\begin{table}
\begin{center}
\begin{tabular}{ccc}
$z_{\rm obs}$ &  $H_{\rm CC}$ [km~s$^{-1}$~Mpc$^{-1}$] &  reference \\
\hline
 & &  \\
0.1 & $69\pm12$ &  \citep{Sim2005} \\
0.17 & $83 \pm 8$  & \citep{Sim2005} \\
0.1791 & $75 \pm 4$  &  \citep{Mor2012} \\
0.1993 & $75 \pm 5$  &  \citep{Mor2012}\\
0.27 & $77 \pm 14$  &  \citep{Sim2005}\\
0.3519 & $83 \pm 14$  &  \citep{Mor2012}\\
0.4 & $95 \pm 17$  &  \citep{Sim2005}\\
0.48 & $97 \pm  60$  &  \citep{Ste2010}\\
0.5929 & $104 \pm 13$  &  \citep{Mor2012}\\
0.6797 & $92 \pm 8$  &  \citep{Mor2012}\\
0.7812 & $105 \pm 12$  &  \citep{Mor2012}\\
0.8754 & $125 \pm 17$  &  \citep{Mor2012}\\
0.88 & $90 \pm 40$  &  \citep{Ste2010}\\
0.9 & $117 \pm 23$  &  \citep{Sim2005}\\
1.037 & $154 \pm 20$  &  \citep{Mor2012}\\
1.3 & $168 \pm 17$  &  \citep{Sim2005}\\
1.363 & $160 \pm 33.6$  &  \citep{Mor2015} \\
1.43 & $177 \pm 18$ &  \citep{Sim2005}\\
1.53 & $140 \pm 14$ &  \citep{Sim2005}\\
1.75 & $202 \pm 40$ &  \citep{Sim2005}\\
1.965 & $186.5 \pm 50.4$  &  \citep{Mor2015} \\
  & &  \\  
\end{tabular}
\caption{Measurements of the Hubble parameter $H_{\rm CC}$ from observations of cosmic chronometers.}
\label{hubpar-cc}
\end{center}
\end{table}

\subsection{Parameters}

The LZ data set constrains redshift remapping, the matter density parameter $\Omega_{\rm m}$, 
the Hubble constant $H_{0}$ and the sound horizon scale $r_{\rm s}$.  As outlined in section 2.1, 
redshift remapping is specified by $6$ parameters with priors arising from fitting cosmological 
models without dark energy and reducing the allowed family solutions to monotonic functions (without distinct features). 
The matter density parameter is a free parameter for fits with open cosmological models and it is fixed at $\Omega_{\rm m}=1$ 
for flat models. Normalization of the Hubble diagram of Type Ia supernovae is an additional nuisance parameter.

Theoretical TT power spectra in our model depends on 8 parameters (or 7 parameters assuming flat models), i.e. 
the cold dark matter density parameter $\Omega_{\rm c}$, the baryonic matter density parameter $\Omega_{\rm b}$, 
the Hubble constant $h$, the scalar spectral index $n_{\rm s}$, normalization of the power spectrum given by $\sigma_{8}$, 
reionization optical depth $\tau$, value of redshift remapping given by the ratio $T_{\rm 0\;obs}/T_{0}$ and parameter 
$A_{\rm L}$ scaling the standard lensing potential power spectrum \citep{Cal2008}. We assume the same 
neutrino sector as in a Planck baseline $\Lambda$CDM cosmological model \citep{Pla2015}, i.e. the effective 
number of neutrinos $N_{\rm eft}=3.046$ and a single massive eigenstate with the corresponding density 
parameter $\Omega_{\nu}h^{2}=0.0006$. We adopt $T_{\rm 0\;obs}=2.7255$~K as our fiducial value of the 
measured CMB temperature \citep{Fix2009}. Finally, we use a default value of the mass fraction in Helium 
assumed in standard Planck cosmological fits, i.e. $Y_{\rm p}=0.2477$.

\section{Results}
In this section we summarize results of fitting dark-matter-dominated cosmological models with redshift remapping. 
We consider two background cosmological models: open cold-dark-matter-dominated models (oCDM$\alpha$) and 
a flat model with $\Omega_{\rm m}=1$ (the Einstein-deSitter model, hereafter EdS$\alpha$). 
The models are constrained by the two data sets described in the previous section:
low-redshift (LZ) data set including Type Ia supernovae, BAO, cosmic chronometers and the local determination of 
the Hubble constant; and high-redshift (HZ) data set including the CMB 
temperature power spectrum from the Planck observations (high multipoles). We present constraints from both data sets 
analyzed separately and in a joint analysis combining both of them (LZ+HZ). We also consider a cosmological fit 
for a subset of the LZ data set which excludes the cosmic chronometer data (LZ-CC).

\subsection{Open model (oCDM$\alpha$)}

\begin{figure*}
\centering
\includegraphics[width=1.00\textwidth]{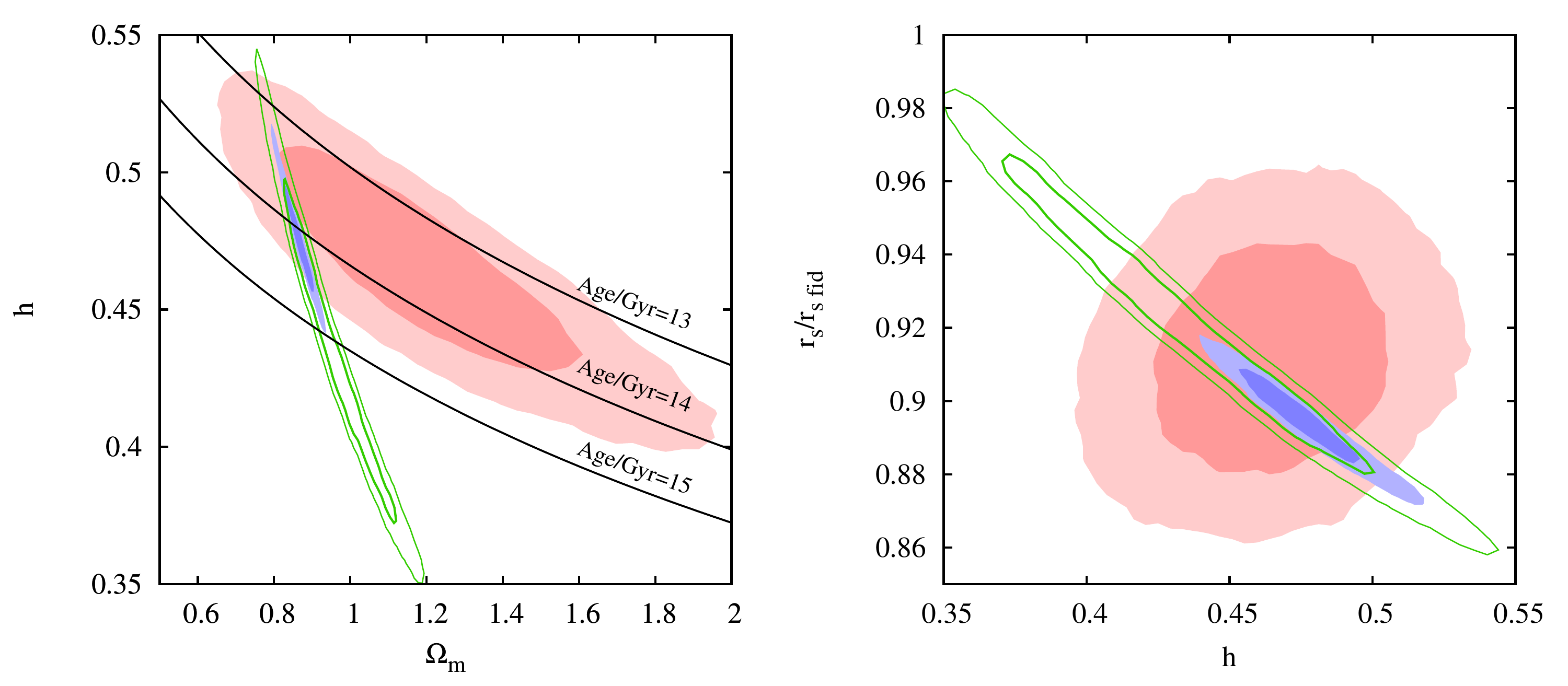}
\caption{Marginalized 68\% and 95\% confidence level constraints on the matter density parameter $\Omega_{\rm m}$, 
the Hubble constant $h$ and the sound horizon scale $r_{\rm s}$ (relative to the fiducial value $r_{\rm s\;fid}=147.78$~Mpc 
for the Planck $\Lambda$CDM cosmology) assuming an open CDM cosmological model with redshift remapping 
(oCDM$\alpha$). The red and green contours show constraints from the $LZ$ and $HZ$ data sets, 
whereas the blue contours from a joint analysis combining both data sets. The black lines are curves of constant age of 
the Universe fixed at 13, 14 and 15~Gyr.}
\label{Om-h-BAO-open}
\end{figure*}

Fig.~\ref{Om-h-BAO-open} shows constraints on three parameters in common to modeling both LZ and HZ data sets: 
the matter density parameter $\Omega_{\rm m}$, the sound horizon scale $r_{\rm s}$ and the Hubble constant $h$. The 
plots serve as an important test of consistency between both data sets. A remarkable agreement between constraints 
from the LZ and HZ data sets demonstrates that redshift remapping lays a self-consistent framework for a joint analysis of the 
CMB and low-redshift probes. The new framework yields consistency between the CMB and low-redshift data 
in terms of constraints on the Hubble constant and the sound horizon scale. As recently shown in \citep{Ber2016}, 
the standard cosmological framework based on a $\Lambda$CDM cosmological model and the standard $a-z_{\rm obs}$ 
fails to pass this consistency test unless the cosmological constant is replaced by a phantom-like dark energy \citep{val2016}.

Both LZ and HZ data sets favor nearly flat cosmological models with a low value of the Hubble constant. The matter 
density parameter from the LZ data set is strongly degenerated with the Hubble constant along the line of constant 
age of the Universe. As shown in Fig.~\ref{Om-h-BAO-open}, the 68\% contour lies between the isochrone lines 
of $13$~Gyr and $14$~Gyr. The corresponding degeneracy line of constraints from the HZ data set is substantially 
steeper than curves of constant age. The age of $14$~Gyr corresponds to $\Omega_{\rm m}\approx0.9$ and 
flatness necessitates the age larger than $15$~Gyr.

Table~\ref{par-open} lists constraints on all parameters. Three columns show the results from separate analyses 
of the two data sets (LZ, HZ) and from a joint analysis combining both of them (LZ+HZ). Compared to the Planck 
cosmological model, the scalar spectral index $n_{\rm s}$ and the deionization optical depth $\tau$ remain unchanged, 
whereas $\sigma_{\rm 8}$  is $7$\% smaller. Constraints on all remaining parameters undergo significant modification. 
Dark-matter-dominated cosmological model with redshift remapping is characterized by higher density of baryons 
and cold dark matter, but nearly the same baryon-to-dark matter fraction as in Planck cosmology. The sound horizon scale 
is $\sim10$\% smaller than its value in the standard $\Lambda$CDM model. A joint analysis of the LZ and HZ data sets 
yields a small curvature with $\Omega_{\rm K}=1-\Omega_{\rm m}=0.13\pm{0.03}$. Finally, the new oCDM$\alpha$ model 
with redshift remapping requires suppression of smoothing of the power spectrum due to lensing based on the standard 
template of the lensing potential. The best-fit amplitude of the lensing potential is $0.65\pm{0.07}$ (65\% of the standard lensing amplitude).

\begin{table*}
\begin{center}
\begin{tabular}{c|c|c|c}
parameter & HZ &  LZ & LZ+HZ\\
\hline
\hline
 & & & \\
$100\Omega_{\rm b}h^{2}$ & $2.90\pm{0.28}$ & - & $3.12\pm{0.13}$ \\
$\Omega_{\rm c}h^{2}$ & $0.152\pm{0.014}$ & - & $0.163\pm{0.005}$\\
$h$ & $0.439\pm{0.042}$ &  $0.464\pm{0.027}$ & $0.475\pm{0.016}$\\
$n_{\rm s}$ & $0.969\pm{0.008}$ & - & $0.968\pm{0.008}$\\
$\tau$ & $0.106\pm{0.050}$ & - & $0.086\pm{0.045}$\\
$\sigma_{8}$ & $0.772\pm{0.034}$ & - & $0.774\pm{0.035}$\\
$T_{\rm obs\;0}/T_{0}$ & $0.920\pm0.028$ & - & $0.896\pm{0.010}$\\
$A_{\rm L}$ & $0.638\pm{0.070}$ & - & $0.650\pm{0.068}$\\
 & & & \\
 \hline
  & & & \\
 $1+\alpha(0.0)$ & -  & $0.63\pm{0.05}$ & $0.64\pm{0.02}$\\
 $1+\alpha(0.5)$ & -  & $0.71\pm{0.04}$ & $0.69\pm{0.02}$\\
 $1+\alpha(1.0)$ & -  & $0.75\pm{0.05}$ & $0.71\pm{0.02}$\\
 $1+\alpha(2.5)$ & -  & $0.84\pm{0.07}$ & $0.75\pm{0.03}$\\
   $\textrm{d}\alpha/\textrm{d}z_{\rm obs}(0.0)$ & - & $0.141\pm0.076$ & $0.115\pm0.065$ \\
 $\textrm{d}\alpha/\textrm{d}z_{\rm obs}(2.5)$ & - & $0.082\pm0.015$ & $0.044\pm0.012$ \\

  & & & \\
\hline
 & & & \\
$\Omega_{\rm m}$ & $0.953\pm{0.098}$  & $1.21\pm{0.27}$ & $0.868\pm{0.031}$ \\
$r_{\rm s}/{\rm Mpc}$ & $135.9\pm{4.1}$  & $134.8\pm{3.1}$ & $132.4\pm{1.4}$\\
${\rm Age}/{\rm Gyr}$ & $15.13\pm{1.15}$  & $13.59\pm{0.41}$ & $14.13\pm{0.37}$\\
 
\end{tabular}
\end{center}
\caption{Posterior mean and standard deviation for the parameters of an open CDM model with redshift remapping 
(oCDM$\alpha$). The left column shows constraints from the HZ data set (Planck CMB temperature power spectrum), the middle column 
from the LZ data set (SN Ia, BAO, $H_{0}$ and cosmic chronometers) and the right column from a joint analysis combining 
both data sets. The top part of the table contains all relevant parameter for 
calculating the CMB temperature power spectrum, 
the middle part lists constraints on the parameters of redshift remapping at the interpolation knots and the bottom part shows 
selected derived parameters from the CMB analysis.}
\label{par-open}
\end{table*}

\begin{figure*}
\centering
\includegraphics[width=1.00\textwidth]{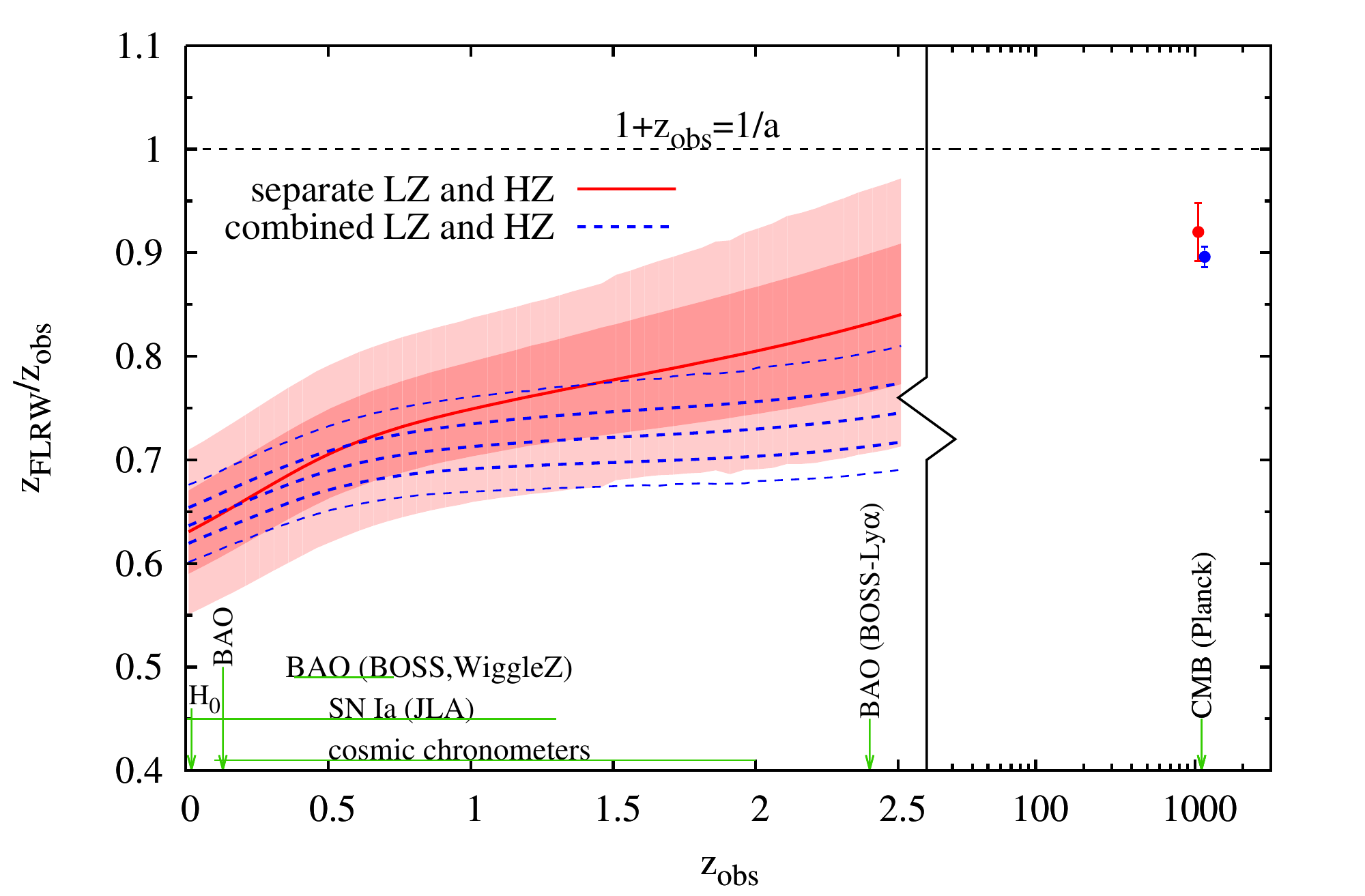}
\caption{Non-parametric constraints on redshift remapping assuming open dark-matter-dominated cosmological models. 
The red and blue contours (data points at CMB redshift) show marginalized 68\% and 95\% confidence level constraints 
from a separate analysis of the LZ (SN Ia, BAO, $H_{0}$ and cosmic chronometers) and HZ (Planck temperature power 
spectrum) data sets, and from a joint analysis combining the two data sets. The green lines 
and arrows show redshift distributions or redshift locations of different cosmological probes in the LZ data set 
($z_{\rm obs}<2.5$) and the HZ data set (CMB). The vertical lines marks the transition between a linear and logarithmic 
scale of $z_{\rm obs}$.}
\label{remapping-open}
\end{figure*}

Fig.~\ref{remapping-open} presents the key result of our work which is a non-parametric reconstruction of redshift 
remapping. The red contours and the red point at $z_{\rm obs}\approx1100$  show constraints from separate analyses 
of the LZ and HZ data sets, whereas the blue from a joint analysis combining both data sets. Without lost of sufficient precision, 
when computing redshift remapping at the CMB redshift, we approximate $(1+z_{\rm FLRW})/(1+z_{\rm obs})=T_{\rm 0\;obs}/T_{0}$ 
(given in Table~\ref{par-open}) by $z_{\rm FLRW}/z_{\rm obs}$. The green lines and arrows show redshift distribution 
(or redshift locations) of the different cosmological probes used in this work.

Although the adopted priors allow for a constant profile, redshift remapping clearly appears to be a function increasing 
with redshift. In particular, the difference between redshift remapping at the ends of the interpolation range, i.e. $z_{\rm obs}=0$ and 
$z_{\rm obs}=2.5$, is $0.21\pm{0.05}$ for LZ and $0.11\pm{0.02}$ 
for LZ+HZ. It is also well visible from Fig.~\ref{remapping-open} that the bulk change of redshift remapping in 
the redshift range of the LZ data set occurs at $z_{\rm obs}\lesssim 0.8$.

Results from independent analyses of the LZ and HZ data sets indicate that redshift remapping may exhibit a plateau 
at $z_{\rm obs}\gtrsim 2.5$ extending up to the recombination redshift. This kind of plateau would be the simplest form of interpolation 
between the LZ data and the CMB, i.e. between $z_{\rm obs}\approx2.5$ and $z_{\rm obs}\approx1100$. Results from the 
joint analysis suggest that redshift remapping should continue to increase at $z_{\rm obs}>2.5$ in order 
to match the value determined from the CMB. This, however, does not rule out the presence of a plateau at higher redshift. 
Assuming a naive linear extrapolation of redshift profile determined from the LZ data set, we can see that redshift 
remapping may reach its value from the CMB at $z_{\rm obs}\approx10$.

Redshift remapping reconstructed from the LZ data is fully consistent with constraints from the joint analysis. 
Adding the HZ data results only in a small reduction of the remapping at $z_{\rm obs}>1$. The remaining part of 
the profile remains nearly unchanged (despite a two fold increase of the reconstruction precision).

The confidence intervals of the reconstructed redshift remapping clearly demonstrate that the data constrain the remapping values 
below 1 at all redshifts. This means that the scale factor assigned to the observed redshift $z_{\rm obs}$ 
in the reconstructed redshift remapping is always larger than that obtained 
from the standard relation $1+z_{\rm obs}=1/a$. From the purely data-based point of view, the deviation of the reconstructed redshift 
remapping from the standard $1+z_{\rm obs}=1/a$ relation corresponds a positive cosmological constant in a $\Lambda$CDM 
cosmological model with the standard mapping between the cosmological redshift and cosmic scale factor. Adding 
a cosmological constant in our background model would result in a strong degeneracy between the expansion history 
and redshift remapping with a one-to-one correspondence between consecutive values of the cosmological constant 
and gradually flattening profiles of redshift remapping, as shown in \citep{Woj2016}.

\begin{figure*}
\centering
\includegraphics[width=1.00\textwidth]{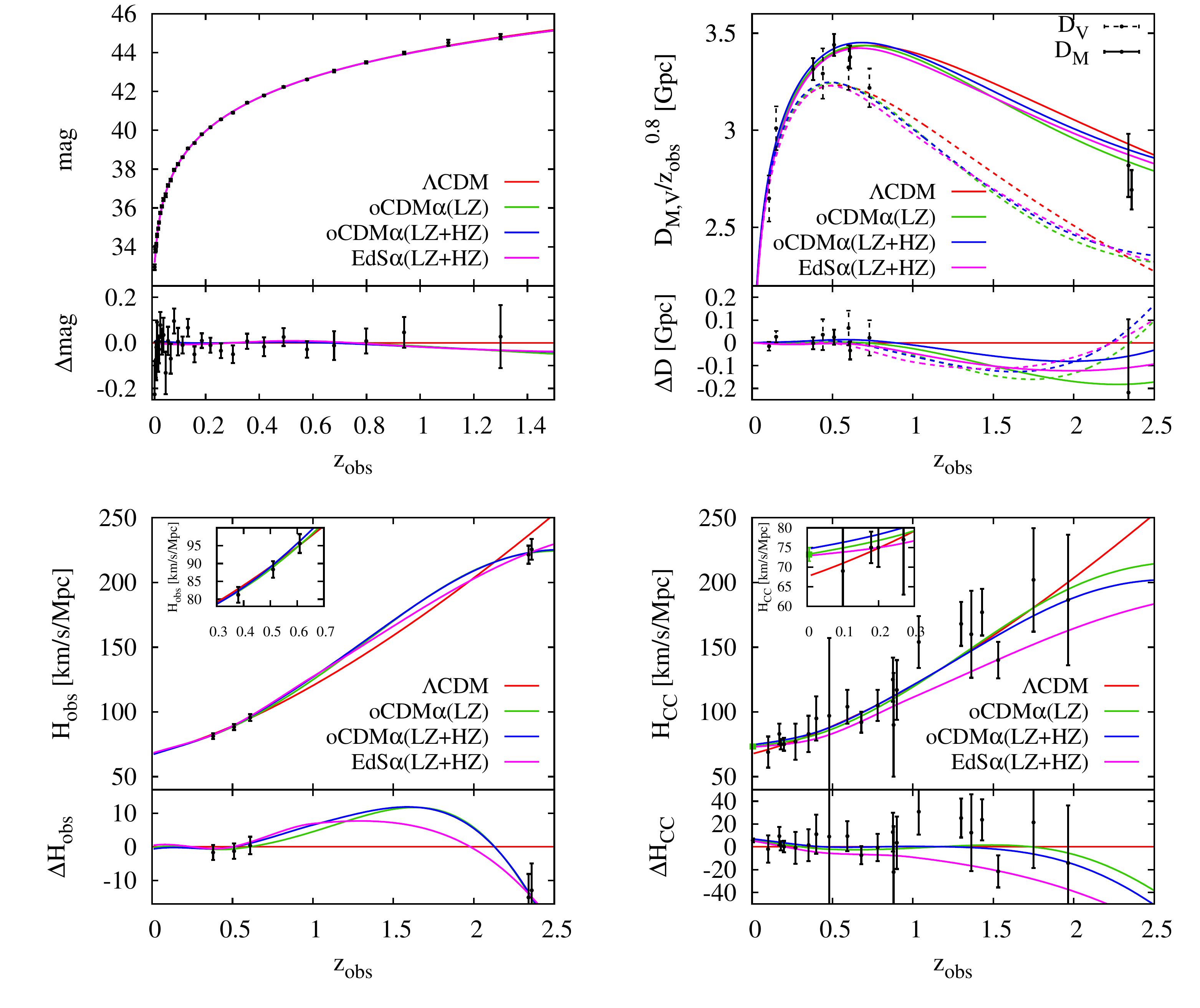}
\caption{The LZ data used in this work compared to best-fitting cosmological models with redshift 
remapping and the Planck $\Lambda$CDM model. From top left clockwise, the panels show apparent 
magnitudes of Type Ia supernovae, distances from BAO observations (6dF, SDSS-MGS, WiggleZ, SDSS-III/BOSS), 
the Hubble parameter $H_{\rm cc}$ estimated from cosmic chronometers and the Hubble parameter $H_{\rm obs}$ 
measured from BAO observations (BOSS). The local determination of the Hubble constant is shown in the 
bottom right panel. Best-fitting models include open CDM (oCDM$\alpha$) and flat CDM (EdS$\alpha$) models 
with redshift remapping, constrained by the LZ data set alone (LZ) and by combined LZ and HZ data (LZ+HZ). 
The red curves show the predictions of the Planck flat $\Lambda$CDM cosmological model (with the standard mapping 
between the cosmological redshift and cosmic scale factor). All residuals are computed relative to the assumed 
Planck cosmological model.
}
\label{res-LZ}
\end{figure*}

In summary, open CDM models with redshift remapping provide excellent fits to all data. Fig.~\ref{res-LZ} compares 
all cosmological measurements in the LZ data set to our best-fitting models and the fiducial flat $\Lambda$CDM cosmological 
model with Planck parameters (and the standard relation between the cosmological redshift and cosmic scale factor). 
Since the comparison of the Hubble parameter from BAO observations involves recalibration due to 
the new value of the sound horizon scale, we include the local measurement of the Hubble constant in the panel 
showing the data of cosmic chronometers. We also combine all constraints on distances from BAO observation in a single panel. 
The solid and dashed lines differentiate between the comoving distance $D_{\rm M}$ and the volume averaged distance 
$D_{\rm V}$.

Our best-fitting models outperform the Planck $\Lambda$CDM cosmological model (with the standard $a-z_{\rm obs}$ mapping) 
in two respects. First, they alleviate a tension between 
Planck observations and the BAO measurements from the Ly-$\alpha$ forest of high redshift quasars. 
The improvement is particularly well visible for the Hubble parameter. The supposed persistence of this tension reported in our previous work \citep{Woj2016} turns out to be a result of assuming a fixed shape of redshift remapping. Second, redshift remapping also eliminates 
a discrepancy between the local measurement of the Hubble constant and its counterpart inferred from Planck data under 
the assumption of a flat $\Lambda$CDM cosmology (see the inset panel for the cosmic chronometer data).

\begin{figure*}
\centering
\includegraphics[width=1.0\textwidth]{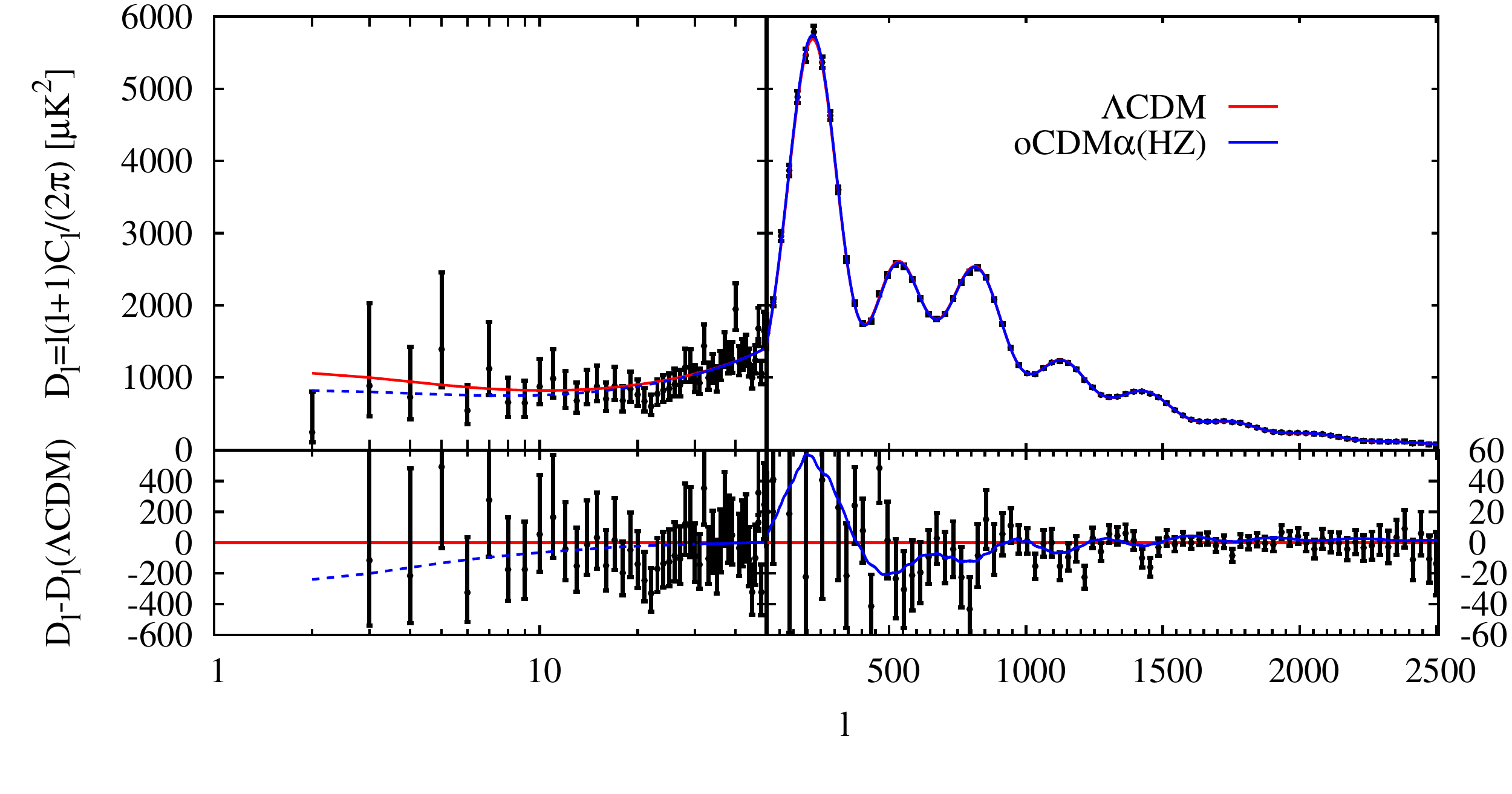}
\caption{Planck CMB temperature power spectrum compared to the best-fitting open CDM model 
with redshift remapping (oCDM$\alpha$, the blue line) and the Planck $\Lambda$CDM cosmology 
($\Lambda$CDM, the red line). Due to small angle approximation, the best-fitting model is constrained by $l\geq30$ multipoles. 
Therefore, its extrapolation to small multiples (the blue dashed line) may not be accurate enough for a detailed 
comparison to the corresponding $\Lambda$CDM model. The displayed Planck data consists of unbinned 
power spectrum at $l<50$ (linear scale on the plot) and binned measurements at $l>50$ 
(logarithmic scale on the plot). Residuals are computed relative to the assumed Planck cosmology. 
The black vertical lines separate a linear and logarithmic scale on the horizontal axis.}
\label{res-HZ}
\end{figure*}

Equally well goodness of fit is achieved for the CMB temperature power spectrum. Fig.~\ref{res-HZ} compares the 
Planck power spectrum to our best-fitting open cosmological model with redshift remapping constrained by 
the HZ data set alone and the Planck $\Lambda$CDM cosmology. Due to small angle approximation, our fit makes 
use only of high multipole part of the power spectrum with $l\geq30$. Therefore, its extrapolation to small multipoles 
(the blue dashed line) may be insufficiently accurate for a detailed comparison to the corresponding $\Lambda$CDM 
power spectrum.

The best-fitting power spectrum of an open model with redshift remapping is essentially indistinguishable from 
its $\Lambda$CDM counterpart. The maximum likelihood value is $-379.8$ compared to $-380.7$ for the 
Planck flat $\Lambda$CDM model with the standard mapping between the cosmological redshift and cosmic 
scale factor ($T_{\rm 0}=T_{\rm 0\;obs}$).

\subsection{Flat model (EdS$\alpha$)}

\begin{figure}
\centering
\includegraphics[width=0.49\textwidth]{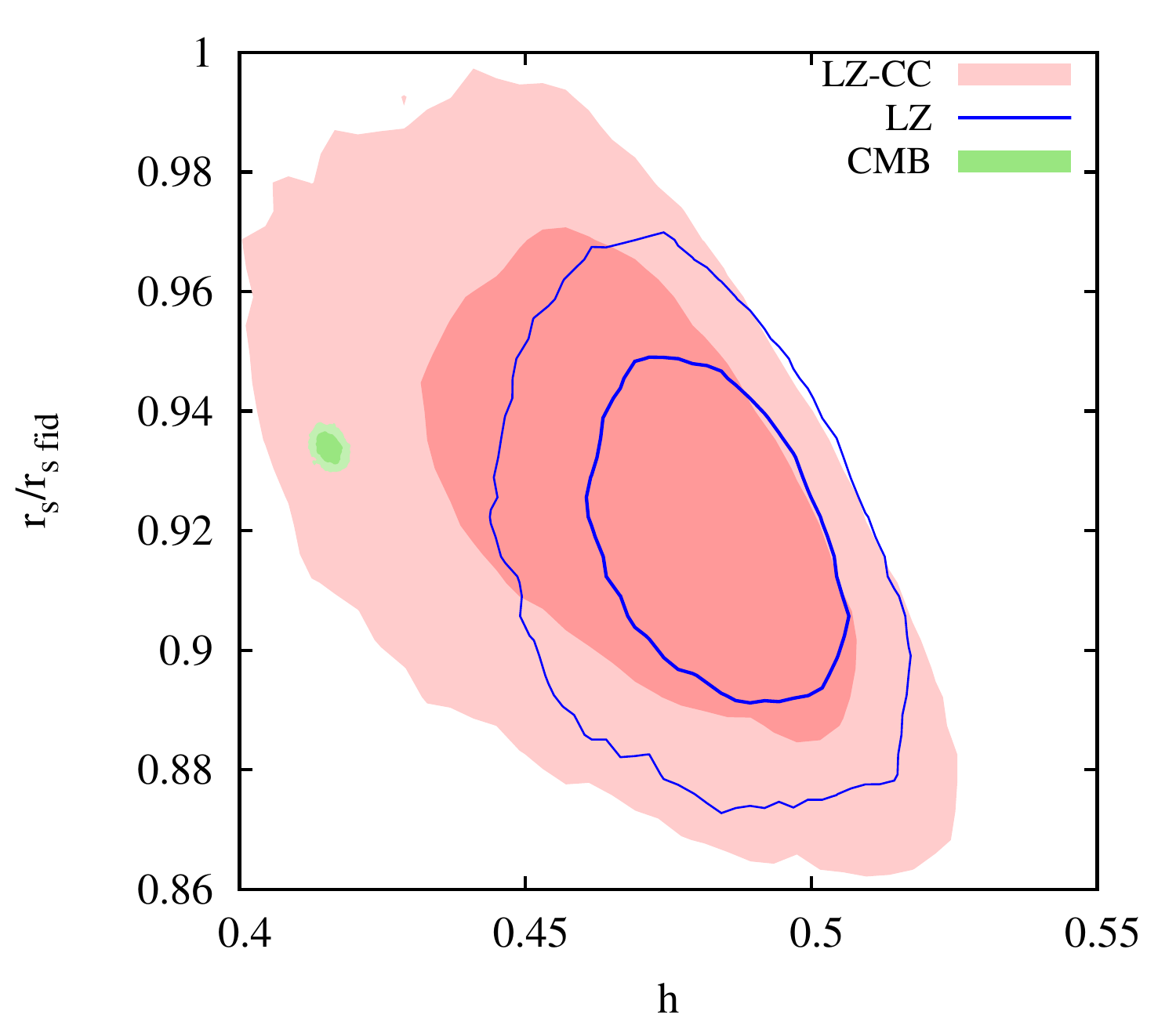}
\caption{Marginalized 68\% and 95\% confidence level constraints on the Hubble constant $h$ and the sound horizon scale 
$r_{\rm s}$ (relative to the fiducial value $r_{\rm s\;fid}=147.78$~Mpc for the Planck $\Lambda$CDM cosmology) assuming 
a flat CDM model (Einstein-deSitter) with redshift remapping (EdS$\alpha$). The blue and green contours show constraints 
from the $LZ$ and $HZ$ data sets, respectively. The red contours show results from an analysis of the the LZ set omitting 
the cosmic chronometer data (LZ-CC).}
\label{BAO-h-EdS}
\end{figure}

As shown in Fig.~\ref{Om-h-BAO-open}, both LZ and HZ data sets, when analyzed separately, are fully consistent with 
a flat CDM model. It is therefore instructive to scrutinize whether or under what conditions the Einstein-deSitter model with 
redshift remapping can provide a satisfactory description of the two combined data sets. The major difficulty arises here 
from the fact the the CMB data constrain the Hubble constant to significantly smaller values than the LZ data set when 
assuming a flat model. This is well visible both in Fig.~\ref{Om-h-BAO-open} and Fig.~\ref{BAO-h-EdS} which shows the marginalized constraints on the Hubble constant and the sound horizon scale for an EdS model with redshift remapping. The latter 
also proves that the problem of a tension between high and low redshift data does not concern the sound horizon scale: 
both the CMB and the LZ data provide fully consistent constraints on $r_{\rm s}$.

Since the degeneracy between $\Omega_{\rm m}$ and $h$ is much stronger for the CMB than for the LZ data 
(see Fig.~\ref{Om-h-BAO-open}), the assumption of flatness results in a much larger increase of the constraining 
power from the CMB than from low-redshift probes. The difference between constraining power of the two data sets 
is also reflected by the difference in size of the contours in Fig.~\ref{BAO-h-EdS}. Bearing in mind the leading role 
of the CMB in a joint fit, we expect that the comparison between the LZ data and the best-fitting joint-fit model 
may reveal which fraction of the LZ data set is responsible for the tension between LZ and HZ data sets in 
terms of constraints on the Hubble constant $h$.

\begin{table*}
\begin{center}
\begin{tabular}{c|c|c|c}
parameter & HZ & LZ-CC & LZ-CC+HZ\\
\hline
\hline
 & & & \\
$100\Omega_{\rm b}h^{2}$ & $2.730\pm{0.057}$  & - & $2.730\pm{0.057}$ \\
$\Omega_{\rm c}h^{2}$ & $0.1449\pm{0.0017}$  & - & $0.1448\pm{0.0017}$\\
$h$ & $0.416\pm{0.002}$   & $0.466\pm{0.026}$ & $0.416\pm{0.002}$\\
$n_{\rm s}$ & $0.966\pm{0.008}$  & - & $0.967\pm{0.008}$\\
$\tau$ & $0.099\pm{0.051}$  & - & $0.100\pm{0.049}$\\
$\sigma_{8}$ & $0.762\pm{0.038}$  & - & $0.763\pm{0.038}$\\
$T_{\rm obs\;0}/T_{0}$ & $0.936\pm0.004$  & - & $0.935\pm0.004$\\
$A_{\rm L}$ & $0.642\pm{0.072}$  & -&  $0.643\pm{0.073}$ok \\
 & & & \\
\hline
 & & & \\
 $1+\alpha(0.0)$ & - & $0.64\pm0.03$ & $0.570\pm0.008$ \\
  $1+\alpha(0.5)$ & - & $0.71\pm0.04$ & $0.617\pm0.007$ \\
   $1+\alpha(1.0)$ & - & $0.75\pm0.05$ & $0.64\pm0.01$ \\
    $1+\alpha(2.5)$ & - & $0.83\pm0.08$ & $0.67\pm0.02$ \\
  $\textrm{d}\alpha/\textrm{d}z_{\rm obs}(0.0)$ & - & $0.152\pm0.081$ & $0.124\pm0.066$ \\
 $\textrm{d}\alpha/\textrm{d}z_{\rm obs}(2.5)$ & - & $0.067\pm0.023$ & $0.024\pm0.010$ \\
  & & & \\
 \hline
 & & & \\
$\Omega_{\rm m}$ & $1$  & $1$ & $1$ \\
$r_{\rm s}/{\rm Mpc}$ & $138.0\pm{0.45}$  & $137.0\pm{4.0}$ & $138.0\pm{0.38}$\\
${\rm Age}/(10^{9}{\rm yrs})$ & $15.68\pm{0.07}$  & $14.0\pm{0.8}$ & $15.68\pm{0.07}$\\
 & \\
\end{tabular}
\end{center}
\caption{Posterior mean and standard deviation for the parameters of a flat CDM model ($\Omega_{\rm m}=1$) with 
redshift remapping. The left column shows constraints from the HZ data set (Planck CMB temperature power spectrum), 
the middle columns from the HZ-CC data set (SN Ia, BAO and $H_{0}$, excluding cosmic chronometers) and 
the right column from a joint analysis combining both data sets. The top part of the table contains all relevant 
parameter for calculating the CMB temperature power spectrum, 
the middle part lists constraints on the parameters of redshift remapping at the interpolation knots and the bottom part shows selected 
derived parameters from the CMB analysis.}
\label{par-EdS}
\end{table*}

The best-fitting model (EdS$\alpha$) from a join analysis combining the LZ and HZ data is shown in Fig.~\ref{res-LZ}. The model exhibits an excellent agreement with all data points but the Hubble parameters from cosmic 
chronometers, in particular at high redshifts $z_{\rm obs}>1$. 
Reanalysis of the LZ set excluding the cosmic chronometer data 
confirms that cosmic chronometers are the main (and probably the only) source of the tension between the Hubble 
constant from the LZ and HZ data (see Fig.~\ref{BAO-h-EdS}). 
Omitting the cosmic chronometer data seems to be the only way to combine constraints from the LZ and HZ sets 
in a fully consistent way under the assumption of a flat cosmological model and this is how we proceed in the following 
part of this section. Without any attempt to justify our choice in a thorough way, we emphasize that the cosmic chronometer 
data set is the most affected by possible astrophysical biases among all cosmological probes compiled in the LZ  data set. 
Keeping in mind the attractive simplicity of the EdS model and an appealing consistency with other LZ probes, we 
think that omitting the cosmic chronometer data is an acceptable tradeoff.

Table~\ref{par-EdS} summarizes constraints on all relevant parameters from separate analyses of the LZ-CC and HZ data sets, 
and from a joint analysis combining both data sets. The assumption of flatness does not change the constraints from the previous 
section for the following parameters: $n_{\rm s}$, $\tau$, $\sigma_{8}$ and $A_{\rm L}$. Likewise an open model, 
the baryon-to-matter fraction is nearly the same as in the Planck $\Lambda$CDM cosmology. The main difference lies in 
substantially smaller value of the Hubble constant and larger age of the Universe, both measured down to a $0.5$\% precision 
in this case (5 fold increase with respect to fits with an open CDM model).

\begin{figure*}
\centering
\includegraphics[width=1.0\textwidth]{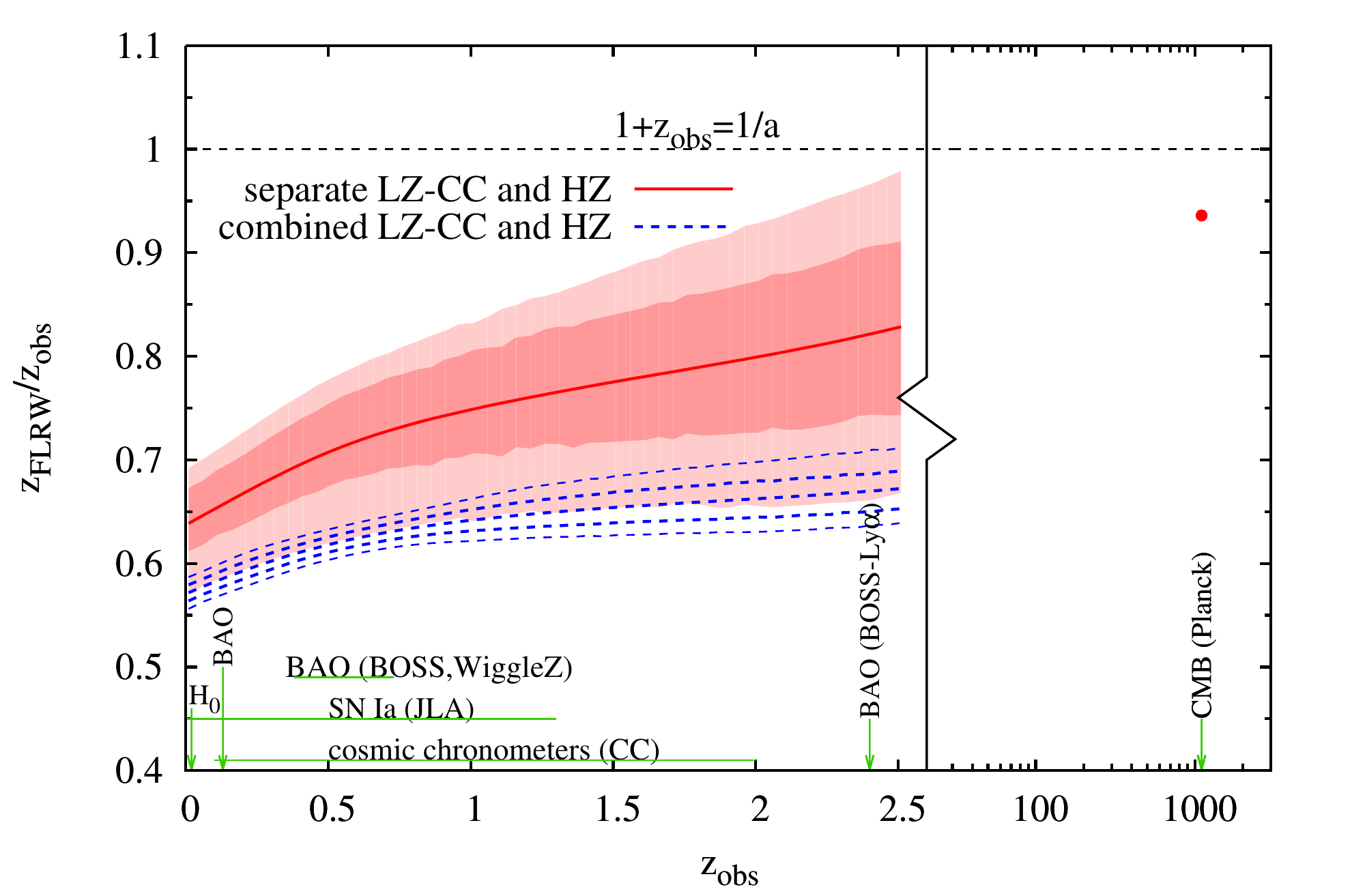}
\caption{Non-parametric constraints on redshift remapping assuming a flat dark-matter-dominated cosmological model 
(the Einstein-deSitter model, EdS$\alpha$). The red and blue contours (and the data point at the redshift of recombination) 
show marginalized 68\% and 95\% confidence level constraints from a separate analysis of the LZ-CC 
(SN Ia, BAO and $H_{\rm 0}$, excluding cosmic chronometers) and 
HZ (Planck temperature power spectrum) data sets, and from a joint analysis combining both data sets. The green lines 
and arrows show redshift distributions or redshift locations of different cosmological probes in the LZ data set 
($z_{\rm obs}<2.5$) and the HZ data set (CMB). The vertical lines marks the transition between a linear and logarithmic 
scale of $z_{\rm obs}$.}
\label{remapping-EdS}
\end{figure*}

Fig.~\ref{remapping-EdS} shows constraints on redshift remapping for the Einstein-deSitter model. The remapping 
reconstructed from the LZ-CC data appears to resemble closely its oCDM$\alpha$ counterpart shown in 
Fig.~\ref{remapping-open}. 
The difference occurs when the CMB data are included in the analysis. In this case, the reconstructed redshift remapping 
appears to have noticeably smaller values 
than in the other reconstruction cases considered in this work. Its redshift profile follows closely the lower $2\sigma$ envelope 
of the remapping reconstructed from the LZ data, regardless if the cosmic chronometer data are included in the fit or not. 
The overall slope, however, is similar to the remapping reconstructed from a joint analysis with an open cosmological model 
(see the blue contours in Fig.~\ref{remapping-open}).

\section{Discussion and conclusions}
We have demonstrated in this work that fitting dark-matter-dominated cosmological model with redshift remapping 
alters several properties and cosmological parameters which are well-established in the standard $\Lambda$CDM 
framework. In this section, we describe and summarize these modifications and explore possible consequences. 
We also enumerate several features of the new models which remain unchanged with respect 
to the Planck cosmology. We elaborate the importance of some of them in the context of sustaining 
consistency with other branches of cosmology such as Big Bang nucleosynthesis.

\subsection{Baryon content}
The baryon fraction in our models is fully compatible with its analogous measurement within the standard 
$\Lambda$CDM framework. Our constraints on $\Omega_{\rm b}/\Omega_{\rm c}$ are $0.191\pm{0.007}$ 
for oCDM$\alpha$(LZ+HZ) and $0.189\pm0.006$ for EdS$\alpha$(LZ-CC+HZ), compared to 
$0.185\pm{0.003}$ derived from the Planck cosmological model. Consistency with the standard baryon fraction 
implies that the new models are not expected to introduce significant modifications to our current knowledge 
on the baryonic content in galaxy clusters. However, complete understanding of possible impact of the new 
models on the predicted properties of galaxy clusters will require carrying out hydrodynamical simulations with new cosmological parameters.

Another important aspect of the baryonic content in cosmological models is the baryon-to-photon number density ratio, 
$\eta=n_{\rm b}/n_{\gamma}$. Redshift remapping modifies the ratio not only through a different baryon density 
parameter, but also through the fact that the number of photons is given by $T_{\rm 0}^{3}$ rather than by 
$T_{\rm obs\;0}^{3}$ (since it is the former which is proportional to $a^{-3}$). Combining these two effects 
leads to the following relation between the new ratio $\eta$ and its value $\eta_{\rm Planck}$ in the Planck 
cosmological model with $\Omega_{\rm b}h^{2}=0.022$:
\begin{equation}
\frac{\eta}{\eta_{\rm Planck}}=\frac{\Omega_{\rm b}h^{2}}{0.022}\Big(\frac{T_{\rm obs\;0}}{T_{0}}\Big)^{3}.
\end{equation}
Using best-fitting models established in a joint analysis of LZ and HZ data sets, 
we find $\eta/\eta_{\rm Planck}=1.020\pm0.014$ for 
the oCDM$\alpha$ model and $\eta/\eta_{\rm Planck}=1.016\pm0.013$ for the EdS$\alpha$ model. 
We obtain virtually no change in $\eta$ compared to the Planck cosmology, what implies that our models 
preserve all standard prediction of the primordial 
nucleosynthesis. Needless to say, the assumed fraction of Helium in our analysis of the CMB power 
spectrum is fully consistent with the resulting baryon-to-photon number density ratio.

\subsection{Hubble constant}
The Hubble constant $H_{0}$ in all models considered in this work appears to be much smaller than its 
locally observed value $H_{\rm obs\;0}$. The ratio of these two values of the Hubble constant is given by 
the local value of redshift remapping, as shown by equation~\ref{hub-loc-obs}, leading to
\begin{equation}\label{H1}
	(H_{\rm 0\;obs}-H_{0})/H_{0} = \left\{
	\begin{array}{ll}
	0.60\pm0.10 & \textrm{(LZ)} \\
	0.575\pm0.046 & \textrm{(LZ+HZ)}
\end{array} \right.
\end{equation}
for the oCDM$\alpha$ model and
\begin{equation}\label{H2}
	(H_{\rm 0\;obs}-H_{0})/H_{0} = \left\{
	\begin{array}{ll}
	0.57\pm0.08 & \textrm{(LZ-CC)} \\
	0.755\pm0.025 & \textrm{(LZ-CC+HZ)}
\end{array} \right.
\end{equation}
for the EdS$\alpha$ model.

The above relative differences between $H_{\rm 0\;obs}$ and $H_{0}$ coincide with the maximum expansion 
rates (relative to the Hubble constant $H_{0}$) in voids formed in the corresponding cosmological models. 
Using our constraints on $\Omega_{\rm m}$, approximating the maximum expansion rate by 
$(3/2)\Omega_{\rm m}^{0.6}$ (see \citep{Ber1997}) and taking its $1/3$ contribution to redshift along 
the light paths, we find
\begin{equation}\label{H1}
	\frac{1}{2}\Omega_{\rm m}^{0.6} = \left\{
	\begin{array}{ll}
	0.56\pm0.07 & \textrm{(LZ)} \\
	0.46\pm0.01 & \textrm{(LZ+HZ)}
\end{array} \right.
\end{equation}
for the oCDM$\alpha$ model and $0.5$ for the EdS$\alpha$ model. Apart from obvious disagreement for the EdS$\alpha$ 
model constrained by the LZ and HZ data sets, the maximum expansion rates are fairly consistent with the corresponding 
relative differences between the local and global values of the Hubble constant. This may be an incidental match, but it 
may also point to a role of voids 
in searching for the physical explanation of redshift remapping. The likelihood of the incident becomes even smaller if 
we realize that the new Hubble constant adjusts the age of the Universe in dark-matter-dominated models so that 
it is fully consistent with the current lower limits determined from observations of globular clusters \citep{Grat2003} 
and metal-poor stars \citep{Bon2013}.

The difference between $H_{\rm 0}$ and $H_{\rm 0\;obs}$ have several important consequences for the interpretation of 
the available observational data. In the following, we describe two interesting examples.

The very first argument for a low value of the total matter density parameter in the Universe was based on 
the measurement of the mass-to-light ratio in galaxy clusters. When combined with the mean luminosity density, 
it lead to $\Omega_{\rm m}\approx0.2$ (see e.g. \citep{Car1997}). Correction of this estimate required by 
adopting models with redshift remapping can be deduced by considering basic scaling relations between the 
Hubble constant and different observables. On one hand, dynamical mass $M\sim h_{\rm obs}^{-1}$ and 
luminosity $L\sim h_{\rm obs}^{-2}$ imply that the mass-to-light ratio $(M/L)\sim h_{\rm obs}$. On the other 
hand, the luminosity density $(L/V)\sim h_{\rm obs}$ in the unit of the critical density $\rho_{\rm c}\sim h^{2}$ 
is $(L/V/\rho_{\rm c})\sim h_{\rm obs}/h^{2}$. Combining these two relations leads to 
$\Omega_{\rm m}\sim (h_{\rm obs}/h)^{2}$ and $\Omega_{\rm m}\approx(0.5-0.6)$ for our constraints 
on $H_{\rm 0\;obs}/H_{0}$. Therefore, the difference between the global and locally measured value of 
the Hubble constant predicted by our models implies the matter density parameter which is compatible 
with dark-matter-dominated cosmological model. Assuming the same bias of this measurement as 
in the standard $\Lambda$CDM model ($\Omega_{\rm m}\approx0.2$ compared to $\Omega_{\rm m}=0.31$ from Planck), 
we find $\Omega_{\rm m}\approx(0.8-0.9)$, in agreement with constraints from analysis of LZ and HZ data.

The second example concerns the mass function. Having shown that constraints on the shape of the power 
spectrum are comparable to the $\Lambda$CDM case, we expect that the main effect modifying the mass function 
in our models stems from a higher value of the matter density parameter. Cosmological models with 
$\Omega_{\rm m}\approx 1$ predict in general higher density of dark matter haloes than observations. Comparing 
the mass functions of the EdS and $\Lambda$CDM model, we find that the former overestimates the abundance 
of haloes by $\sim (0.5-0.6)$dex (see \citep{Jen2001}). Needless to say, this statement assumes implicitly 
equality between $h$ and $h_{\rm obs}$ that is not true in the framework with redshift remapping. 
Correct comparison of the new mass function with observations requires applying conversion 
of the units from $\hMpc$ to $h_{\rm obs}^{-1}$~Mpc. This conversion lowers the mass function 
by $3\log_{10}(h_{\rm obs}/h)\approx(0.6-0.7)$ what happens to compensate in large part the effect of large value of 
$\Omega_{\rm m}$. Although any conclusion that the mass function in our models can be reconciled with the observed 
abundance of galaxy clusters would be premature, we emphasize that the above estimate shows that this 
situation is totally conceivable. This problem will be addressed in more detail in our ongoing work.

The idea that a nearly Einstein-de-Sitter cosmological model with a low value of the Hubble constant 
has the potential to fit observations in not new \citep[see e.g.][]{Bar1995}. However, our studies clearly demonstrate 
that reconciling this model with the highest precision cosmological observations to date requires modification 
of the standard mapping between the cosmological redshift and cosmic scale factor.

An alternative way to reconcile an Einstein-de-Sitter model with observations was recently proposed 
by \citet{Racz2016} who showed that as low value of the Hubble constant as $h=0.4$ can be amplified to the observed 
value $h_{\rm obs}=0.73$ when the standard FLRW expansion is replaced by the local expansion rate averaged over 
cosmic space. The similarity between this approach and redshift remapping in terms of amplifying the Hubble constant makes 
us think whether there is a deeper connection between both models. Future work shall shed more light on 
this issue.

\subsection{Lensing}
\begin{figure}
\centering
\includegraphics[width=0.49\textwidth]{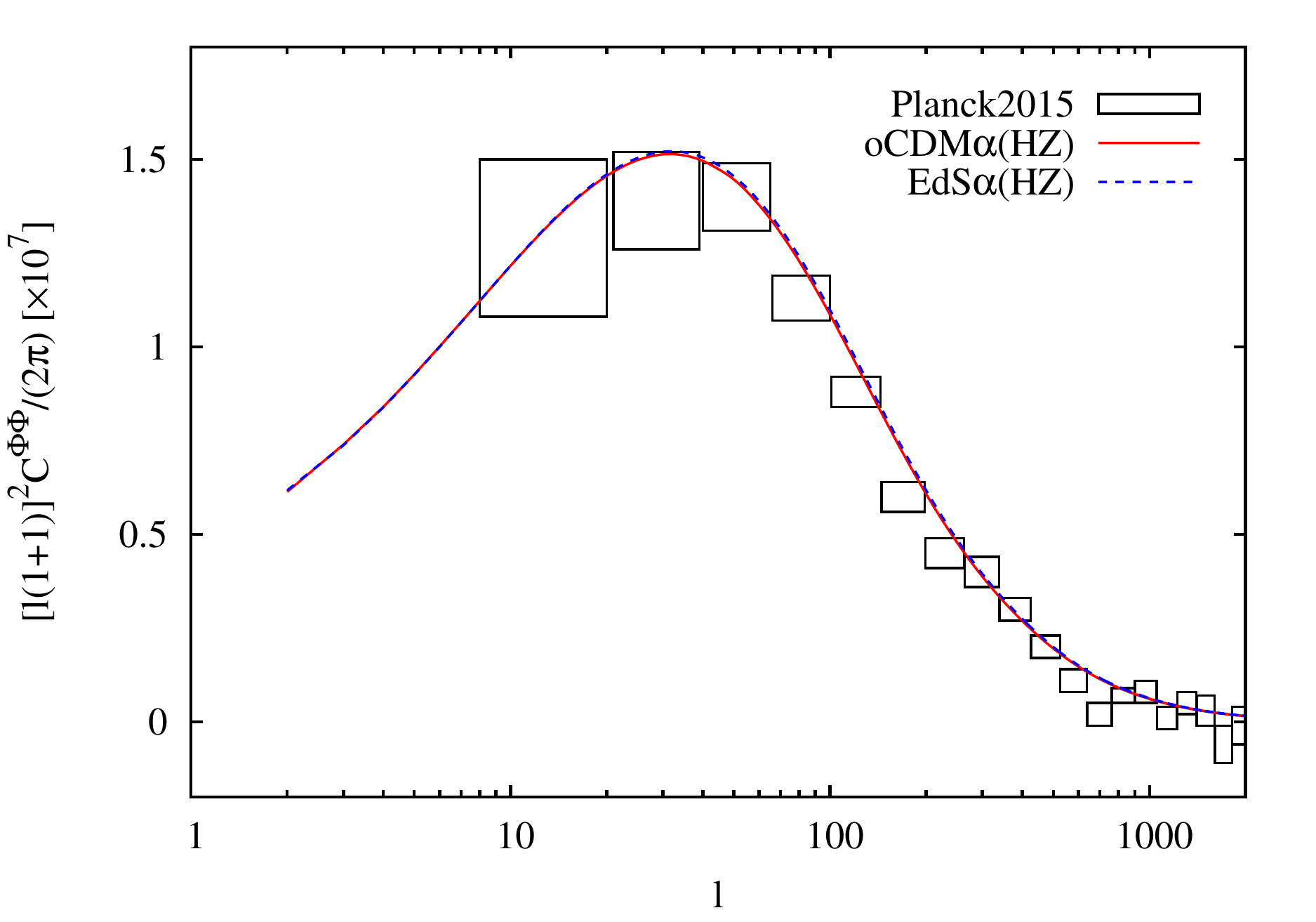}
\caption{Lensing potential power spectrum measured by Planck compared to best-fitting models with redshift remapping and 
open (oCDM$\alpha$) or flat (EdS$\alpha$) cosmological model. Both models are constrained by the CMB temperature power 
spectrum from Planck observations (HZ data set).}
\label{lensing-potential}
\end{figure}

Both open and flat model with redshift remapping require the suppression of the standard lensing potential 
to fit the temperature power spectrum. In both cases, the amplitude of the lensing potential is $35$\% smaller 
at a $5\sigma$ confidence level than the standard amplitude computed for the corresponding cosmological 
models. As a sanity check, in Fig.~\ref{lensing-potential} we compare the lensing power spectrum measured 
from Planck observations \citep{Planck2015b} to the power spectrum of best-fitting models constrained by 
the HZ data (with $A_{L}\approx 0.64$) discussed in our work. 
Since the power spectrum reflects information integrated along the line of sight and thus it is independent of 
redshift, scaling the multipoles according to redshift remapping is not applicable here and the power spectrum 
is identical to the output of \textit{camb}.

Fig.~\ref{lensing-potential} demonstrates an excellent agreement between power spectra of best-fitting models 
and the measurement. It is also straightforward to imagine that using the standard lensing amplitude ($A_{\rm L}=1$) 
would result in a strong tension with the data. Therefore, the suppression of the lensing amplitude appears to be 
a relevant condition for sustainability of our dark-matter-dominated cosmological fits with redshift remapping. 
It seems natural to suspect that the suppression of the lensing potential is coupled somehow 
with redshift remapping. This coupling may be a useful guide in searching for a physical model of redshift remapping.

\subsection{Age}
Observational constraints on dark-matter-dominated models with redshift remapping result in the age of the Universe 
which does not fall below the lower limits determined 
from observations of globular clusters and isolated metal-poor stars \citep{Grat2003,Bon2013}. 
Relatively long ages in models with $\Omega_{\rm m}\approx 1$ are attained due to low values of the Hubble constant. 
It shall be regarded as an interesting coincidence that the measured values of $\Omega_{\rm m}$ and $H_{\rm 0}$ 
in our models always imply the age no smaller than $\sim14$~Gyrs. For open models, the age is consistent with its estimate 
in the Planck cosmological model (see Table~\ref{par-open}), but for the EdS model the CMB requires the age larger 
by $1.3$~Gyrs at high confidence level (see Table~\ref{par-EdS}).

\begin{figure*}
\centering
\includegraphics[width=1.0\textwidth]{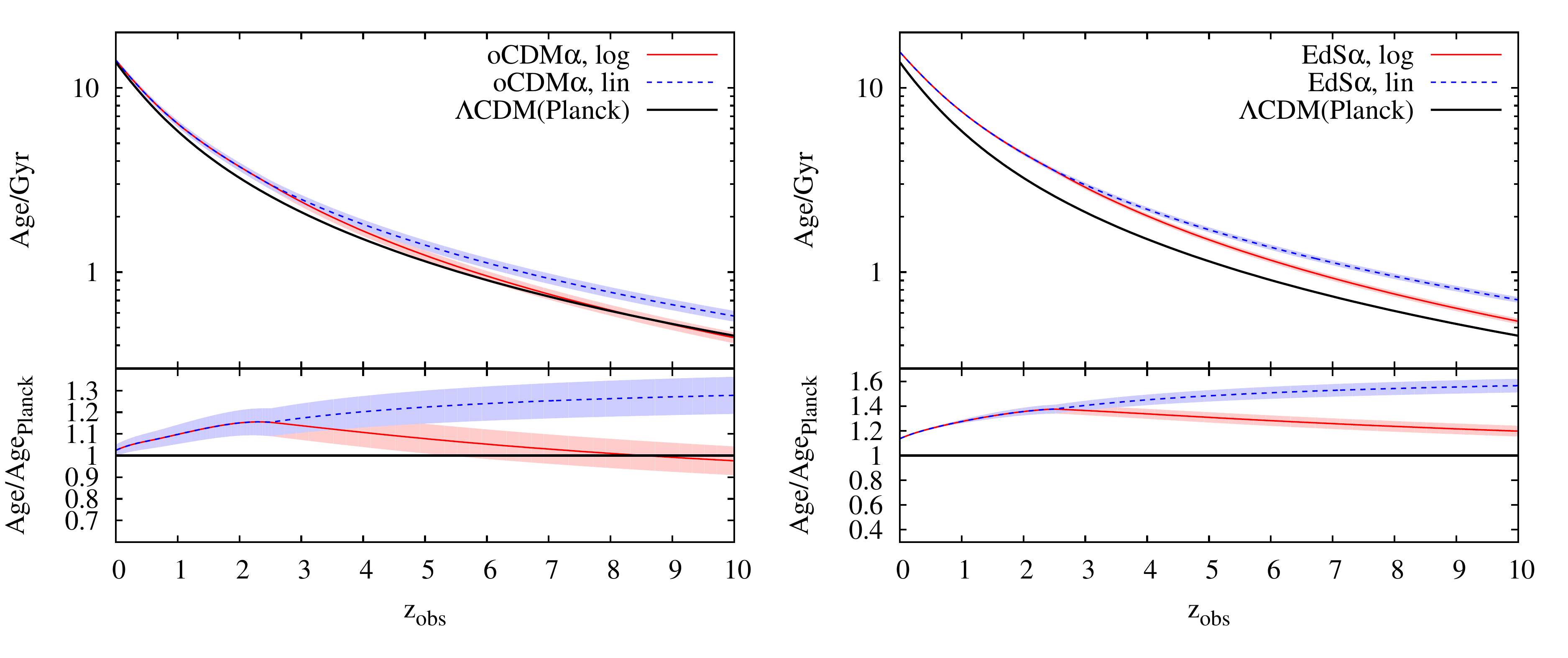}
\caption{The age of the Universe as a function of the observed redshift $z_{\rm obs}$ in the open (oCDM$\alpha$, left panel) and 
flat (EdS$\alpha$, right panel) dark-matter-dominated cosmological with redshift remapping. The red and blue lines show results 
based on logarithmic and linear interpolation between constraints on redshift remapping from the LZ data 
($z_{\rm obs}\leq 2.5$) and the HZ data set ($z_{\rm obs}\approx 1100$), respectively. The bottom panels show 
the ratio of the age in the new models to the corresponding age in the Planck $\Lambda$CDM cosmology.}
\label{age}
\end{figure*}

The new models modify the age of the Universe at all redshifts. Fig.~\ref{age} shows the age of the Universe as a function of 
the observed redshift. The age is computed using the standard formula incorporating redshift remapping, i.e.
\begin{equation}
t=\int_{0}^{z_{\rm obs}[1+\alpha(z_{\rm obs})]}\frac{\textrm{d}z_{\rm FLRW}}{(1+z_{\rm FLRW})H(z_{\rm FLRW})}
\end{equation}
For redshifts larger than $z_{\rm obs}=2.5$ (the upper limit of redshift remapping reconstructed 
from the LZ data set), we assume a simple interpolation between redshift remapping at $z_{\rm obs}=2.5$ and 
$z_{\rm obs}\approx1100$ (constrained by the CMB). We consider linear and logarithmic interpolations.

For the open model with redshift remapping (left panel in Fig.~\ref{age}), the current age of the Universe 
is nearly the same as in the Planck cosmology. However, the new model predicts older Universe at high redshifts 
than Planck. In particular, the Universe is 10\% older at $z_{\rm obs}=2.5$ and, depending on the actual redshift remapping 
between the current LZ data set and the CMB, it can be 30\% older at redshift $z\sim10$. For the EdS model with 
redshift remapping, the Universe is even older, e.g. 15\% older at the present time and up to 60\% older 
at $z_{\rm obs}\sim 10$ than in the Planck cosmology.

This general trend of higher estimates of the age in new models with redshift remapping is very interesting in the context 
of observations of surprisingly well-evolved objects at high redshift such as quasars with supermassive black holes or 
galaxies with large amount of dust. The existence of such objects poses a challenge to theoretical models describing 
formation of dust \citep{Gal2011}, supermassive black holes \citep{Mor2011} or first stars and early galaxies in general. Since all 
well-formed high-redshift objects require unusually rapid formation, it is natural to consider the possibility that 
this is a global anomaly of the standard $\Lambda$CDM cosmological model predicting relatively too strong 
compression of time at high redshift \citep{Mel2014}. Following this line of reasoning, we notice that our models 
with redshift remapping have the potential to alleviate this problem, as clearly demonstrated in Fig.~\ref{age}. 

\subsection{Redshift remapping}

The observationally reconstructed redshift remapping increases with the observed redshift. There is also a strong signature 
that the remapping reaches a plateau at redshift much smaller than the recombination. Using simple extrapolation of redshift 
remapping reconstructed within the redshift range of the LZ data set, we find that this plateau begins at $z_{\rm obs}\approx(5-10)$ 
for the open model.

Our approach has a status of a phenomenological model and thus we do not intend to address the problem what 
is a possible physical cause of the reconstructed redshift remapping. There is doubt, however, that the existence of 
plateau described above implies that any potential physical mechanism should operate only at late epoch of 
cosmic evolution, although it affects both light from low-redshift objects and the CMB photons. It is tempting to associate 
this epoch with the growth on highly non-linear structures. In addition, coincidence between the maximum expansion 
rate in voids and the local value of redshift remapping (which determines the ratio of the actual to locally measured 
Hubble constant) suggest that the non-linear structures which may play a relevant role here are voids.

Typical models explaining the cosmic acceleration as an effect of a superHubble expansion involves voids whose 
size exceed substantially the scale of inhomogeneities limited by the BAO. This apparent conflict 
between the models and observations seems to result from adopting an oversimplified model of void topology. 
According to recent studies \citep{Fal2015,Ram2016}, void as a region without non-linear structures percolates nearly 
all parts of cosmic space and fills 90-95\% of its volume. Such void is infinite (despite a finite scale of the density fluctuations) 
and, in a way, it functions as a medium for propagation of light in cosmic space. 
If non-linear evolution generates strong effects on the actual metric, in particular local boosts 
of the expansion rate demonstrated in recent fully relativistic simulations 
\citep{Ben2016,Gib2016}, void can effectively change the standard mapping between the cosmological redshift 
and cosmic scale factor. The problem whether the resulting redshift remapping can reproduce our non-parametric 
reconstruction from observations can be addressed in future studies combining recently introduced relativistic 
cosmological simulations with the novel concept of persistently percolating void.

\subsection{Cosmological distance scales}

\begin{figure}
\centering
\includegraphics[width=0.49\textwidth]{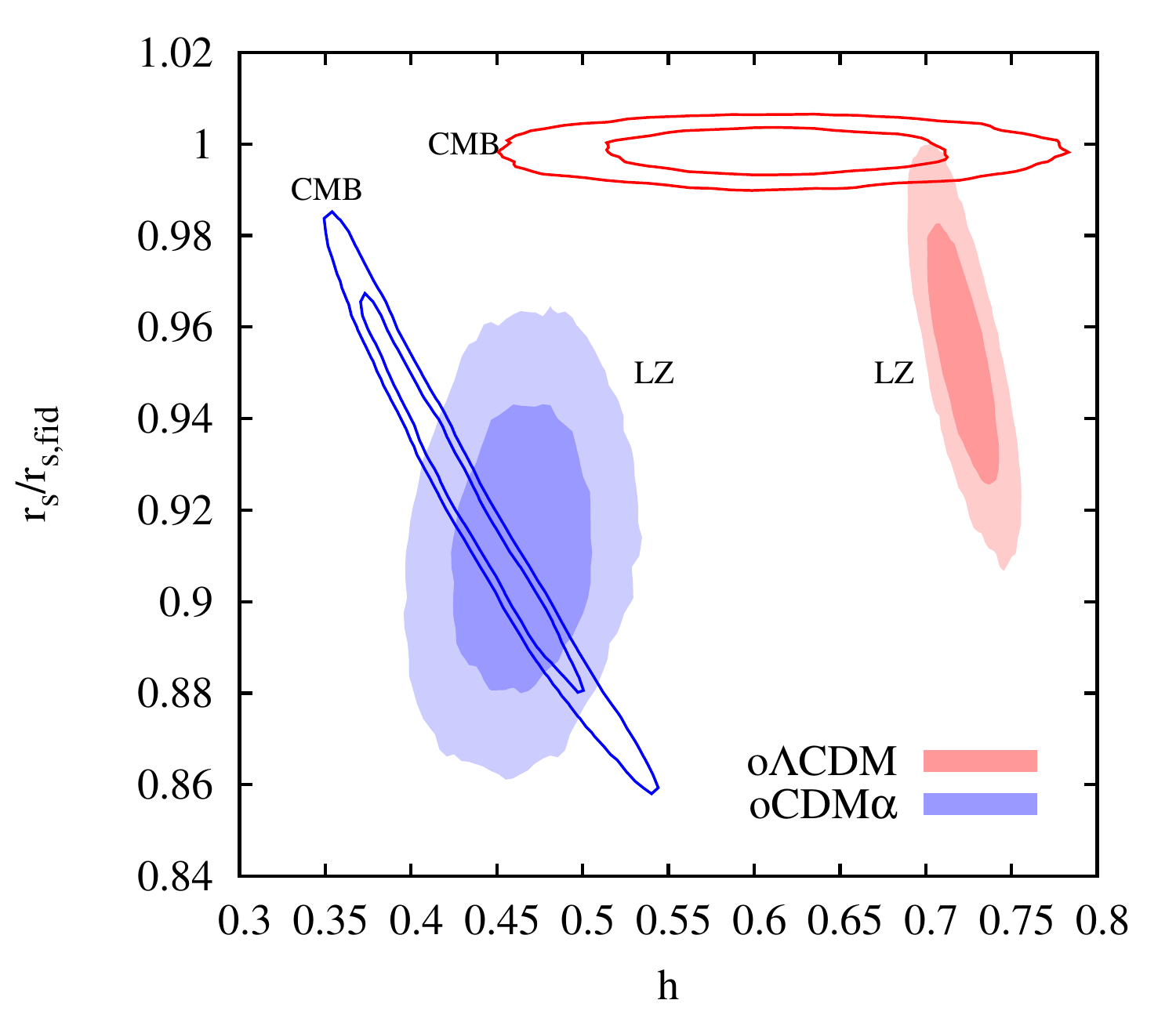}
\caption{The Hubble constant and the BAO scale at low (LZ data set) and high redshifts (HZ data set: CMB) 
for two cosmological models: open $\Lambda$CDM model and open CDM model with redshift remapping. The 
new CDM model with redshift remapping resolves the discrepancy between the distance scales at low and 
high redshifts in the standard $\Lambda$CDM model.}
\label{distance-cal}
\end{figure}

The new class of models with observationally reconstructed redshift remapping have in general higher degrees of freedom. 
Therefore, it is not surprising that certain anomalies apparent in the standard $\Lambda$CDM model, in particular 
the BAO measurements from the Ly-$\alpha$ forest of BOSS quasars, are absorbed by the new models. On the other hand, the 
consistency between parameters inferred independently from the LZ and HZ data sets should be regarded as a highly 
non-trivial result. Of particular importance is the consistency between the LZ and HZ observations in terms of the Hubble 
constant and the BAO scale, two fundamental parameters setting the absolute scales of cosmological distances.

Fig.~\ref{distance-cal} compares constraints on the Hubble constant and the BAO scale from the LZ and HZ 
observations assuming two cosmological models: open $\Lambda$CDM model and open CDM model with redshift 
remapping. It is clearly visible that replacing the cosmological constant by redshift remapping eliminates the tension 
between the distance scales at low and high redshifts, apparent in the standard $\Lambda$CDM model. Alleviating 
this tension without any modification of the standard mapping between the cosmological redshift and cosmic scale 
factor requires introduction of a dynamical model of dark energy \citep{val2016}.

\section{Summary}
We made use of all primary cosmological probes (the CMB temperature power spectrum from Planck observations, 
Type Ia supernovae, BAO, cosmic chronometers and the local determination of the Hubble constant) 
to reconstruct a relation between the cosmological redshift $z_{\rm obs}$ and cosmic scale factor $a$ (redshift remapping) 
and to determine parameters of the assumed dark-matter-dominated cosmological model. Our main findings can be 
summarized as follows.
\begin{itemize}
\item The new cosmological model with redshift remapping provide excellent fits to all data. It eliminates two well-recognized tensions within the 
standard $\Lambda$CDM cosmological model: anomalous values of the BAO signal determined from the Ly-$\alpha$ forest of BOSS quasars 
and discrepancy between the local and CMB-based determinations of the Hubble constant. All observational signatures of cosmic acceleration are accounted for by the reconstructed redshift remapping instead of dark energy.

\item A joint analysis of the CMB temperature power spectrum and low-redshift probes results in a nearly flat cosmological model 
with $\Omega_{\rm m}=0.87\pm{0.03}$. When omitting the cosmic chronometer data, the data are consistent with the Einstein-deSitter model.

\item Constraints on the Hubble constant and the BAO scale from low-redshift data are fully consistent with those from the CMB temperature power spectrum (respectively 40\% and 10\% smaller than the standard values in the Planck $\Lambda$CDM 
cosmological model).

\item The difference between redshift remapping reconstructed from observations and the standard $1+z_{\rm obs}=1/a$ 
relation decreases monotonically with the observed redshift. The reconstructed redshift remapping exhibits the largest 
deviation from the standard $a-z_{\rm obs}$ relation at $z_{\rm obs}=0$ and a plausible constant deviation at $z_{\rm obs}\gtrsim 10$. 
Both features point to the fact that any possible physical mechanism behind redshift remapping should operate at late times of cosmic evolution.

\item The reconstructed redshift remapping predicts a significant difference between the actual and locally measured values of the 
Hubble constant. This difference coincides with the maximum expansion rate in cosmic voids formed in the corresponding cosmological model with $\Omega_{\rm m}\approx 0.9$. The actual Hubble constant has sufficiently low value to compensate effects of high dark matter density parameter. In particular, the age of the Universe is $\sim14$~Gyrs and is fully compatible with the lower limits determined 
from astrophysical estimates. Simple interpolations between redshift remapping constrained by the low-redshift probes and the CMB redshift show, however, that the Universe may be up to 30~\% older at high redshifts.

\item The new cosmological model has a comparable shape of the power spectrum as the standard $\Lambda$CDM model. It is nearly indistinguishable from the Planck cosmology in terms of the baryon fraction and the photon-to-baryon number density ratio. The latter 
guarantees that the newly established model is compatible with the predictions of Big Bang nucleosynthesis.

\item Fitting the CMB temperature power spectrum requires the suppression of the standard lensing template by $\sim35$~\%, both for 
the open and flat cosmological model. The corrected lensing template is fully consistent with independent measurement of the lensing 
power spectrum from Planck observations.

\subsection{Prospects for future tests}

The most stringent tests of our model in foreseeable future will rely on increasing the constraining power of low-redshift probes. 
The primary role will be played by the BAO measurements from the DESI \citep{desi2013} and EUCLID \citep{euc2011} surveys. The future 
observations of the BAO signal will improve the current status of this probe in several respects: higher precision, filling the current redshift 
gap at $0.8<z_{\rm obs}<2.2$, verification of the BAO anomaly from the Ly-$\alpha$ forest of BOSS quasars and extending the redshift 
coverage up to $z_{\rm obs}=3.5$ (thanks to the DESI Ly-$\alpha$ forest observations). Precise BAO measurements at high redshifts will be of 
particular importance, because this is where our model predicts deviations from the standard $\Lambda$CDM model (see Fig.~\ref{res-LZ}).

One of the key features of our model is its capability of reconciling the local and CMB-based measurements of the Hubble constant (discrepant 
when interpreted in the framework of the standard $\Lambda$CDM model). From this point of view, future corroboration of the current 
determination of the Hubble constant in a low-redshift universe will be essential. The best strategy will arguably rely on complementing the 
measurements based on Type Ia supernovae with other techniques, for example observations of gravitational lens time delays. The accuracy 
and the precision of the latter method have been substantially improved and the current determinations of the Hubble constant 
strengthen the tension between the measurements of this parameter in the local and high-redshift universe \citep{Bon2017}. The future 
progress of this measurement will be facilitated by the LSST survey \citep{lsst2012} as well as many independent dedicated observational 
campaigns.

There is no doubt that enabling our model to predict certain subtle observables going beyond basic cosmography will require 
understanding physics behind redshift remapping. For example, calculating redshift drift \citep[tiny change of the observed redshifts at the rate 
$\sim10^{-10}$ over a period of $10$ years; see][]{San1962,Loe1998} involves evaluating time derivative of redshift remapping what goes beyond predictions of our 
model at its current phenomenological level. The currently unknown contribution from redshift remapping to redshift drift is the first term in 
the following expression:
\begin{equation}
\frac{\textrm{d}z_{\rm obs}}{{\rm d}t}=-z_{\rm obs}\frac{{\rm d}\ln(1+\alpha)}{{\rm d}t}+\frac{1}{1+\alpha}\frac{{\rm d}z_{\rm FLRW}}{{\rm d}t}.
\end{equation}
Redshift drift will be measured in foreseeable future by the Extremely Large Telescope (ELT) at redshifts $2<z_{\rm obs}<5$ \citep{Lis2008} and quite 
likely by the Square Kilometer Array (SKA)  in a complimentary redshift range $0<z_{\rm obs}<1$ \citep{Abd2015}. Bearing in mind that 
measuring redshift drift is arguably considered the ultimate test of the standard cosmological model, it becomes clear how relevant is 
to raise the status of cosmological models with redshift remapping to the level of complete physical models.

\end{itemize}

\section*{Acknowledgments}
The authors thank an anonymous referee for the constructive comments and suggestions. RW acknowledges support through 
the Porat Postdoctoral Fellowship. RW is grateful for the hospitality of Dark Cosmology Centre where a part of this study 
was accomplished. FP thanks the support from the MINECO grant AYA2014-60641-C2-1-P.

\bibliography{bibGravz}

\appendix
\section{Effect of binning}
\begin{figure}
\centering
\includegraphics[width=0.50\textwidth]{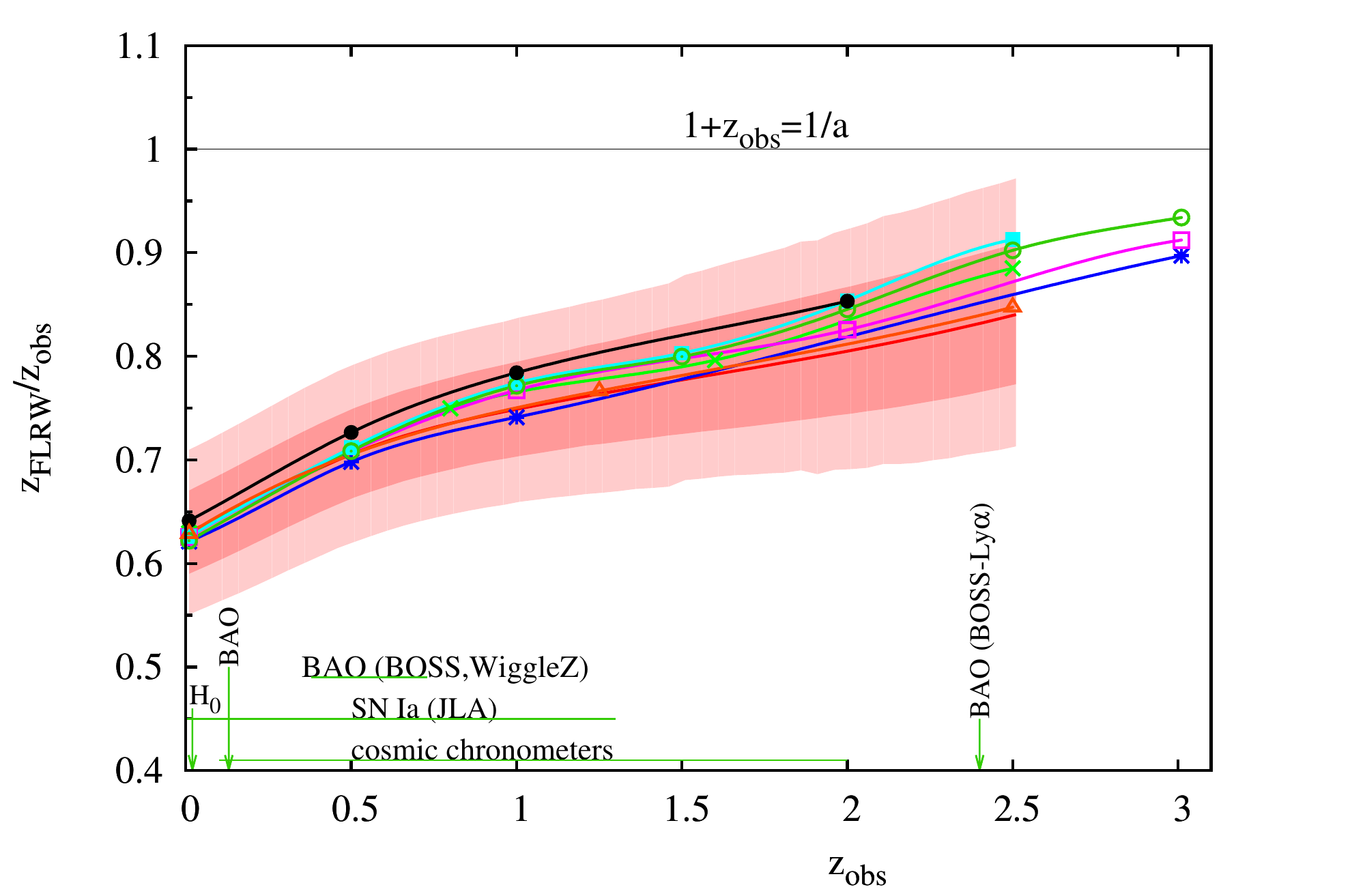}
\caption{Effect of binning defining the knots of the interpolation on best-fitting profile of redshift remapping inferred from the LZ data. The red 
contours show marginalized 68\% and 95\% confidence level constraints from our main analysis. Solid lines show best-fitting profiles obtained 
for several different versions of binning and the symbols indicate the positions of the interpolation knots. All profiles lie well within the 68\% confidence 
level envelope of the main results.}
\label{alpha-profile-bins}
\end{figure}

Here we demonstrate that our choice of binning adopted for the interpolation do not affect robustness of our results. 
Fig.~\ref{alpha-profile-bins} shows best-fitting profiles obtained for several different ways of binning (with different bin sizes, 
number of bins and the position of the last bin). All profiles lie well within the error envelope of the main constraints and 
the cases with an increased number of bins do not reveal any additional features of redshift remapping.

\end{document}